\documentclass[11pt]{article}
\newcommand{\Comment}[1]{{}}
\usepackage{amsfonts,amsthm,amsmath,amssymb,slashed}
\usepackage[textwidth = 430 pt, textheight = 630 pt]{geometry}
\usepackage{color}

\Comment{\usepackage{color}
\definecolor{MyDarkBlue}{rgb}{0.15,0.15,0.45}
\usepackage[linktocpage=true]{hyperref}
\hypersetup{
colorlinks=true,
citecolor=MyDarkBlue,
linkcolor=MyDarkBlue,
urlcolor=MyDarkBlue,
pdfauthor={Daniel W.F. Alves, Carlos Hoyos, Horatiu Nastase and Jacob Sonnenschein},
pdftitle={The title},
pdfsubject={hep-th}
}

\usepackage[numbers,sort&compress]{natbib}
\usepackage{hypernat}}
\usepackage{graphicx}
\usepackage{cite}

\newcommand\ignore[1]{}
\def\one{{\,\hbox{1\kern-.8mm l}}}

\def\Tr{{\rm Tr\, }}

\def\a{\alpha}\def\b{\beta}

\def\d{\partial}

\def\Tr{\mathop{\rm Tr}\nolimits}

\newcommand{\Cset}{{\,\,{{{^{_{\pmb{\mid}}}}\kern-.45em{\mathrm C}}}}}

\newcommand{\be}{\begin{equation}}
\newcommand{\bea}{\begin{eqnarray}}

\newcommand{\ee}{\end{equation}}
\newcommand{\eea}{\end{eqnarray}}

\begin{document}

\renewcommand{\thefootnote}{\fnsymbol{footnote}}

\makeatletter
\@addtoreset{equation}{section}
\makeatother
\renewcommand{\theequation}{\thesection.\arabic{equation}}

\rightline{}
\rightline{}




\begin{center}
{\LARGE \bf{\sc Knotted solutions, from electromagnetism to fluid dynamics}}
\end{center} 
 \vspace{1truecm}
\thispagestyle{empty} \centerline{
{\large \bf {\sc Daniel W.F. Alves${}^{a}$}}\footnote{E-mail address: \Comment{\href{mailto:dwfalves@ift.unesp.br}}{\tt dwfalves@ift.unesp.br}},
{\large \bf {\sc Carlos Hoyos${}^{b}$}}\footnote{E-mail address: \Comment{\href{mailto:hoyoscarlos@uniovi.es}}{\tt hoyoscarlos@uniovi.es}},
{\large \bf {\sc Horatiu Nastase${}^{a}$}}\footnote{E-mail address: \Comment{\href{mailto:nastase@ift.unesp.br}}{\tt nastase@ift.unesp.br}}
{\bf{\sc and}}
{\large \bf {\sc Jacob Sonnenschein${}^{c}$}}\footnote{E-mail address: \Comment{\href{mailto:cobi@post.tau.ac.il}}{\tt cobi@post.tau.ac.il}}
                                                        }

\vspace{.5cm}


\centerline{{\it ${}^a$Instituto de F\'{i}sica Te\'{o}rica, UNESP-Universidade Estadual Paulista}} 
\centerline{{\it R. Dr. Bento T. Ferraz 271, Bl. II, Sao Paulo 01140-070, SP, Brazil}}
\vspace{.3cm}
\centerline{{\it ${}^b$Department of Physics, Universidade de Oviedo,}}
\centerline{{\it Avda. Calvo Sotelo 18, 33007, Oviedo, Spain }} 
\vspace{.3cm}
\centerline{{\it ${}^c$School of Physics and Astronomy,}}
\centerline{{\it The Raymond and Beverly Sackler Faculty of Exact Sciences, }} \centerline{{\it Tel Aviv University, Ramat Aviv 69978, Israel}}

\vspace{1truecm}

\thispagestyle{empty}

\centerline{\sc Abstract}

\vspace{.4truecm}

\begin{center}
\begin{minipage}[c]{380pt}
{\noindent Knotted solutions to electromagnetism and fluid dynamics are investigated, based on relations we find between the two subjects. 
We can write fluid dynamics in electromagnetism language, but only on an initial surface, or for linear perturbations, and we use this map 
to find knotted fluid solutions, as well as new electromagnetic solutions. 
We find that knotted solutions of Maxwell electromagnetism are also solutions of more general nonlinear 
theories, like Born-Infeld, and including ones which contain quantum corrections from couplings with other modes, like Euler-Heisenberg
and string theory DBI. Null configurations in electromagnetism can be described as a null pressureless fluid, and from this map 
we can find null fluid knotted solutions. A type of nonrelativistic reduction of the relativistic fluid equations is described, which allows us 
to find also solutions of the (nonrelativistic) Euler's equations.

}
\end{minipage}
\end{center}

\vspace{.5cm}

\setcounter{page}{0}
\setcounter{tocdepth}{2}

\newpage

\renewcommand{\thefootnote}{\arabic{footnote}}
\setcounter{footnote}{0}

\linespread{1.1}
\parskip 4pt



\section{Introduction}

Knotted solutions, often solutions with a nonzero Hopf index, or ``Hopfions", appear in several  areas of physics, though not always as explicit
solutions of the equations of motion. In Maxwell's electromagnetism without sources, despite the theory's simplicity, such solutions were first written 
only in \cite{Ranada:1989wc,ranada1990knotted} by Ranada, after the early work by Trautman in \cite{Trautman:1977im}. 
There are null solutions, namely with $\vec{F}^2=0 $, where  $\vec{F}=\vec{E}+i\vec{B}$ is the Riemann-Sielberstein (RS) vector, and hence 
$\vec E\cdot\vec B= \vec E^2-\vec B^2= 0$, but there are also more general partially null  solutions, arising from different type of constructions (Bateman's vs. 
Ranada's), as we will review in the paper. Electromagnetism is 
a noninteracting theory, so we can linearly superpose solutions,   and therefore  the solutions are not  solitonic  ones. Moreover, the knotted
solutions do not keep a fixed shape, but rather "spread out" as time passes, which distinguishes them from solitons (famously, John Scott Russel saw a 
solitonic wave in a canal in Scotland that kept its shape for a very long time, which was the start of the study of solitons). 
The topological nature of the solutions  is associated with ``helicities" ${\cal H}_{ab}$, $a,b=e\ ( electric)$ or $m\ (magnetic)$, 
written like spatial integrals of Chern-Simons type terms,
$\int d^3x \epsilon^{ijk}A_i\d_j B_k$, that are conserved for the null configurations for which 
$F_{\mu\nu}F^{\mu\nu}=\epsilon^{\mu\nu\rho\sigma} F_{\mu\nu}F_{\rho\sigma}=0$. 
In Bateman's construction the electric and magnetic fields are expressed in terms of two complex scalar
 fields $\a$ and $\b$. In \cite{Kedia:2013bw,besieris2009hopf}
 it was shown that from  any solution $(\a,\b)$  one can easily find families of solutions by applying on them any 
  holomorphic transformations, $f(\a,\b), g(\a,\b)$.
In particular, if one starts from the Hopfion solution, for which the helicities are  ${\cal H}_{ee}={\cal H}_{mm}= 1$, one can 
get  ``$(p,q)-$knotted solutions" by applying the transformations $\a\rightarrow (\a)^p,\b\rightarrow (\b)^q$.
In \cite{Hoyos:2015bxa}, it was found that by starting from topologically trivial null solutions, like one of constant electric and 
magnetic fields perpendicular to each other, one can derive topologically non-trivial solutions like the  Hopfion solution by applying special 
conformal  transformations with {\em complex} rather than real parameters.
By applying other conformal transformations with such parameters one can enlarge the zoo of knotted solutions.  
For a review of this subject, and a more complete list of references, see \cite{arrayas2016knots}.

On the other hand, knotted solutions are also important in fluid dynamics. In fact, after the original work of Helmholtz in 1858, Lord Kelvin 
found the topological robustness of knotted fluid lines and thus started the theory of knots. But this field remained mostly in the domain of 
abstract study of knots and their evolution \cite{kambe1971motion}, where it became a fertile subject (for reviews see 
\cite{ricca1996topological,ricca1998applications,ricca2009new}, see also the book \cite{arnold1999topological}).
Given that fluid dynamics is complicated and nonlinear, it is perhaps not surprising that theoretical solutions of the fluid equations (Euler in the 
case of the ideal fluid) were very scarce, experimental observations of fluid knots were only made recently \cite{kleckner2013creation}, 
and some numerical constructions were obtained in \cite{proment2012vortex}. Moffatt \cite{moffatt1969degree} started the analytical analysis of fluid knots 
by defining a {\em conserved} fluid flow ``helicity" ${\cal H}_v$, similar to ${\cal H}_{mm}$ in electromagnetism, if we replace the fluid velocity 
$\vec{v}$ with the magnetic vector potential $\vec{A}$, and writing some explicit solutions with 
${\cal H}_v\neq 0$, though in very special cases. Their properties were studied further in \cite{moffatt1992helicity,moffatt1992helicity2}.
Other solutions were found in \cite{ValarMorgulis,crowdy_2004,Ifidon2015216}, and initial conditions with nonzero ${\cal H}_v$ were 
shown in \cite{kuznetsov1980topological}. A connection that was much used (for instance, \cite{Kholodenko:2014apa,Kholodenko:2014wfa,Boozer})
is between magnetohydrodynamics (motion of a fluid coupled to 
electromagnetism) and only fluid dynamics. One maps the magnetic field $\vec{B}$ to the vorticity $\vec{\omega}$, but the fluid 
velocity $\vec{v}$ remains $\vec{v}$, which means that one restricts to the space of solutions where $\vec{B}=\vec{\omega}=\vec{\nabla}\times
\vec{v}$. Then it was found that in magnetohydrodynamics one can define a ``velocity of the lines of force" $\vec{v}_p=(\vec{E}\times\vec{H})/H^2$
\cite{Newcomb} and measure it, and under a ``frozen field condition" $\vec{v}_p$ can coincide with the velocity $\vec{v}$ of the fluid transporting it, 
e.g. \cite{Boozer}, and one can transport the electromagnetic Hopfion \cite{Irvine}.

In this paper we extend and elaborate 
on results already presented in the letter \cite{Alves:2017ggb}, using relations between electromagnetism and fluid dynamics to 
find more (time dependent) knotted solutions for both. First, we find that fluid dynamics can be rewritten in electromagnetism language, with 
some restrictions. Either considering linearized solutions, or just restricting to initial conditions, we can map fluid dynamics to electromagnetism, allowing 
us to find some new electromagnetic solutions, and reversely, to construct some fluid dynamics solutions. 
We describe the nonlinearly defined sector of linear electromagnetism using  Bateman's construction.
Moreover, we find that nonlinear theories of electromagnetism, like Born-Infeld or theories obtained
by integrating out quantum degrees of freedom, like Euler-Heisenberg, also have the knotted solutions as solutions. 
Secondly, using the general procedure of writing an interacting quantum system at large distances as a fluid, by an expansion in derivatives, 
we find that electromagnetism can be written as a null pressureless relativistic fluid ($P=0$, $u^\mu u_\mu=0$). 
We can do a sort of nonrelativistic dimensional reduction of the null fluid equations, in lightcone coordinates, for solutions independent of 
$x^+$, and obtain the nonrelativistic Euler equations with pressure, allowing us to find yet other kinds of solutions. 

The paper is organized as follows. In section 2, we review the knotted solutions of electromagnetism, their topological Hopf index, and 
Bateman's and Ranada's constructions for them. We also show in subsection 2.5 that they are solutions of general nonlinear theories of electromagnetism. 
In section 3, we  review 
known knotted solutions of fluid dynamics, and their conserved helicity. In section 4, we write the Euler fluid as a restricted version of 
electromagnetism, and in section 5 we use this map to find more solutions of electromagnetism. 
In section 6, we show that electromagnetism can be written as a null fluid, provided we restrict to the null subspace ($\vec{E}\cdot\vec{B}=0
=\vec{E}^2-\vec{B}^2$). In section 7, we construct new fluid solutions, first by mapping the Bateman knotted solutions of electromagnetism to it, 
then by finding a map from the null fluid to Euler's equations with pressure, and using a dimensional reduction, and finally by 
mapping the Ranada type solutions to the fluid. In section 8 we conclude. 

\section{Knotted solutions in electromagnetism}

In this section we review knotted solutions of electromagnetism, i.e. solutions that have nonzero values for some conserved ``helicities", in 
both Bateman's construction and Ranada's construction. The simplest solution is the ``Hopfion" solution. 
For a more extensive review, see \cite{arrayas2016knots}.

\subsection{Electromagnetism and helicities}

We will be interested in solutions to the free Maxwell electromagnetism, without sources. The Maxwell's equations without sources are, 
in non-relativistic notation,
\bea
\vec{\nabla}\times \vec{E}=-\frac{\d \vec{B}}{\d t}; && \vec{\nabla}\cdot \vec{E}=0\cr
\vec{\nabla}\times \vec{B}=+\frac{1}{c^2}\frac{\d \vec{E}}{\d t}; && \vec{\nabla}\cdot \vec{B}=0.
\eea
From now on, we will put $c=1$, unless otherwise stated.

The fundamental fields are the scalar potential $\phi$ and the vector potential $\vec{A}$, as 
\be
\vec{E}=-\frac{\d}{\d t}\vec{A}-\vec{\nabla}\phi;\;\;\;\;
\vec{B}=\vec{\nabla}\times \vec{A}.
\ee
In the absence of sources, we can work in a gauge with $A_0=\phi=0$. But then we still have a gauge invariance
\be
\delta\vec{A}=\vec{\nabla}\a(\vec{x})\;,
\ee
that leaves invariant the gauge, since $\delta A_0\equiv \delta\phi=\d_0 \a=0$. Note however that we have still $\vec{A}=\vec{A}(\vec{x},t)$.

But moreover, since $\vec{\nabla}\cdot \vec{E}=0$ in the absence of sources, 
we can also introduce another vector field $\vec{C}$ generating $\vec{E}$ in the same way as $\vec{A}$ generates $\vec{B}$. Since all we need is 
$\vec{A}$, $\vec{C}$ is superfluous, but is introduced for symmetry. Then 
\be
\vec{E}=\vec{\nabla}\times \vec{C};\;\;\; \vec{B}=\vec{\nabla}\times \vec{A}\;,
\ee
so $\vec{C}$ is defined by $-\frac{\d}{\d t}\vec{A}\equiv \vec{\nabla}\times \vec{C}$, which amounts to electromagnetic duality
(so $\vec{C}$ is the electromagnetic dual vector potential). Moreover, we also have $\vec{B}=-\d_t \vec{C}$, which means a completely duality
symmetric electromagnetic formulation.

{\bf Conserved ``helicities"}

We can introduce integrals over space defined as Chern-Simons forms, the electric helicity as a Chern-Simons form of $\vec{C}$,
\be
H_{ee}=\int d^3x \vec{C}\cdot \vec{E}=\int d^3x \vec{C}\cdot \vec{\nabla}\times \vec{C}=\int d^3x \epsilon^{ijk}C_i\d_j C_k\;,
\ee
and its electromagnetic dual, the magnetic helicity as a Chern-Simons form of $\vec{A}$,
\be
H_{mm}=\int d^3x \vec{A}\cdot \vec{B}=\int d^3x\epsilon^{ijk}A_i\d_j A_k\;,
\ee
and as BF forms, the cross helicities, the eletromagnetic one, 
\be
H_{em}=\int d^3x \vec{C}\cdot\vec{B}=\int d^3x \epsilon^{ijk}C_i\d_j A_k\;,
\ee
and its electromagnetic dual, the magnetoelectric one,
\be
H_{me}=\int d^3x \vec{A}\cdot\vec{E}=\int d^3x \epsilon^{ijk}A_i\d_k C_k.
\ee
Being Chern-Simons and BF forms, they are gauge invariant under the 3-dimensional gauge transformations generated by $\a(\vec{x})$, in case 
the transformation of the Abelian fields $\vec{A}(\vec{x},t)$ doesn't have any global issues, i.e. is not a large gauge transformation. Under large
gauge transformations, they can change by an integer times $2\pi$. Note however that, since we have a $\vec{A}(\vec{x},t)$, conservation 
of these helicities in time is not guaranteed, so neither is having integer values for them. 

The time variation of these helicities is found to be 
\bea
\d_t H_{mm}&=&\int d^3x (\d_t \vec{A}\cdot \vec{B}+\vec{A}\cdot \d_t\vec{B})=-\int d^3x(\vec{E}\cdot\vec{B}+\vec{A}\cdot(\vec{\nabla}\times\vec{E}))\cr
&=&-2\int d^3x \vec{E}\cdot\vec{B}\cr
\d_t H_{ee}&=&\int d^3x (\d_t \vec{C}\cdot \vec{E}+\vec{C}\cdot \d_t\vec{E})=-\int d^3x(\vec{B}\cdot \vec{E}+\vec{C}\cdot (\vec{\nabla}\times \vec{B}))\cr
&=&-2\int d^3x \vec{E}\cdot \vec{B}\cr
\d_t H_{me}&=& \int d^3x(\d_t \vec{A}\cdot \vec{E}+\vec{A}\cdot\d_t\vec{E})=\int d^3x(-\vec{E}\cdot\vec{E}+\vec{A}\cdot(\vec{\nabla}\times\vec{B}))\cr
&=&-\int d^3x(\vec{E}^2-\vec{B}^2)\;,\cr
\d_t H_{em}&=& \int d^3x(\d_t \vec{C}\cdot \vec{B}+\vec{C}\cdot \d_t \vec{B})=\int d^3x(-\vec{B}\cdot \vec{B}+\vec{C}\cdot (\vec{\nabla}\times \vec{E}))\cr
&=&\int d^3x (\vec{E^2}-\vec{B}^2)\;,
\eea
where we have used Maxwell's equations and partial integration. Thus conservation of $H_{mm}$ and $H_{ee}$ is equivalent to the 
vanishing of $\vec{E}\cdot \vec{B}$, and of $H_{em}$ and $H_{me}$ to the vanishing of $\vec{E^2}-\vec{B}^2$. 

We will treat two cases: when only $\vec{E}\cdot \vec{B}=0$ (so that only $H_{ee}$ and $H_{mm}$ are conserved), and when also 
$\vec{E}^2-\vec{B}^2=0$, so all helicities are conserved.

Moreover, for these configurations, one can find  
``knotted" solutions, characterized by nonzero values for these helicities, but for which the electric and magnetic field configurations have
nonzero linking numbers (topological quantities). 

\subsection{The Hopf map and Hopf index}

The basic knotted solution, the ``Hopfion", is related to the Hopf map, and has a nonzero Hopf index for the (first) Hopf map, $S^3\rightarrow 
S^2=\mathbb{CP}^1$. We will therefore review here the 
relevant concepts and properties of it. One can find some of them for instance in \cite{Nastase:2009zu}, as well as in various mathematics books.

When $S^2$ is understood as $\mathbb{CP}^1$, the resulting map becomes the Hopf fibration: 
consider $Z^1=X^1+iX^2$ and $Z^2=X^3+iX^4$ complex, satisfying $|Z^1|^2+|Z^2|^2=1$. This defines an $S^3$, relation invariant under 
multiplication by a phase, $Z^\a\rightarrow e^{i\theta}Z^\a$, which is the $U(1)$ fiber of the Hopf fibration. The stereographically projected 
complex coordinates on $S^2$ are ($x^i=X^i/X^4, i=1,2,3$)
\be
W=\frac{x_1+ix_2}{1-x_3}=\frac{X^1+iX^2}{X^3+iX^4}=\frac{Z^1}{Z^2}.
\ee
They are invariant under $Z^\a\rightarrow e^{i\theta}Z^\a$, but also under the more general $Z^\a\rightarrow\lambda Z^\a$, with $\lambda$ complex, 
so they define a map from $S^2$ ($W$) to $\mathbb{CP}^1$ ($Z^\a$). 

This allows a simple generalization (embedding) of the Hopf map into $S^5\rightarrow \mathbb{CP}^2$,  $S^7\rightarrow \mathbb{CP}^3$, and 
more generally $S^{2k+1}\rightarrow \mathbb{CP}^k$. More details can be found in \cite{Nastase:2009zu}. 

The usual formulation of the Hopf map is as a map from 
\be
\sum_{\a=1,2}Z^\a Z_\a^\dagger=1
\ee
to 
\be
\sum_{i=1}^3x_i x^i=1\;,
\ee
via
\be
x_i=Z^\dagger_\a{{(\tilde \sigma_i)}^\a}_\b Z^\b \equiv Z^\b {{(\sigma_i)}_\b}^\a Z^\dagger_\a.
\ee
Here $\sigma_i$ are the Pauli matrices and $\tilde \sigma$ their transposed. 

As an aside, 
note that there is also the second Hopf map, $S^7\rightarrow S^4$, related to the quaterionic algebra. Define $x_A$, $A=1,...,5$ Euclidean coordinates
on $S^4$, i.e. $\sum_{A=1}^5 x_A x_A=1$, and $\sum_{\a=1}^4 |Z^\a|^2=1$, then the map is 
\be
x_A=Z^\b{(\Gamma_A)^\a}_\b Z^\dagger_\a\;,
\ee
where $\Gamma_A$ are $SO(5)$ gamma matrices. There is a quaternionic formulation as well. 
There is also a third Hopf map, $S^{15}\rightarrow S^8$, related to the octonionic algebra. Consider {\em real} objects $g^\a$, $\a=1,...,16$, 
with $\sum_\a g^T_\a g^\a=1$, and $x_A$. $A=1,...,9$ with $\sum_A x_A x_A=1$, then the map is 
\be
x_A=g_\a^T(\Gamma_A)^{\a\b} g_\b.
\ee

The Hopf map defined above has Hopf index, i.e. ``winding number" one. 

An explicit solution for the map viewed as $S^3\rightarrow S^2$, where we use the identification $S^3\equiv \mathbb{R}^3\cup \{\infty\}$, parametrized 
by $(x,y,z)$, and $S^2\equiv \mathbb{C}\cup\{\infty\}$, parametrized by $(\phi,\bar \phi)$, is 
\be
\phi_H(x,y,z)=\frac{2(x+iz)}{2z+i(r^2-1)};\;\;\; r^2=x^2+y^2+z^2.\label{phiH}
\ee

{\bf The Hopf index}

The Hopf index can be defined in terms of a (complex) scalar field defined on $\mathbb{R}^3\cup \{\infty\}\simeq S^3$, 
$\phi(x_1,x_2,x_3)$, taking value on $\mathbb{C}\cup\{\infty\}\simeq S^2$. In both cases, the element at infinity can be added because of a 
{\em boundary condition} on the field, that it goes to a constant (or zero) at infinity. 

The complex scalar field $\phi$ defines an area 2-form on $S^2$,
\be
{\cal \omega}=\frac{1}{2\pi i}\frac{d\bar\phi\wedge d\phi}{(1+|\phi|^2)^2}\;,
\ee
whose pullback onto $S^3$, ${\cal F}$, is a 2-form on $S^3$,
\be
{\cal F}=n^*\omega=\frac{1}{2}F_{ij}dx^i\wedge dx^j=\frac{1}{4\pi i}\frac{\d_i \bar\phi\d_j\phi-\d_j\bar\phi\d_i\phi}{(1+|\phi|^2)^2}dx^i\wedge dx^j\label{2form}
\ee

Because the second cohomology group of the 3-sphere is trivial, $H^2(S^3)=0$, the pull-back of the area form is an exact 2-form, 
${\cal F}=d{\cal A}$, so the associated Chern-Simons form on $S^3$ is a topological index (winding number), the Hopf index,
\be
N=\int_{S^3}{\cal A}\wedge {\cal F}.
\ee

The map $\phi(x_1,x_2,x_3)$ can be thought of also as a map $\vec{n}:\mathbb{R}^3\rightarrow S^2$, modified to a map $S^3\rightarrow S^2$ 
by the boundary condition that $\vec{n}$ goes to a constant at infinity, so that a field theory made from it has finite energy. 
Here $\vec{n}^2=1$, so it is defined on $S^2$. 

The map in terms of the field $\vec{n}(x^\mu)$ can moreover be related to a map taking values in 
$SU(2)$, by the relation 
\be
U=e^{i\vec{n}\cdot \vec{\sigma}}\;,
\ee
where $\vec{\sigma}$ are Pauli matrices. 

There is a winding number expressed in terms of $\vec{n}$  as 
\be
N'=\frac{-i}{24\pi^2}\int \epsilon^{ijk}\Tr[(\d_i\vec{n}\cdot \vec{\sigma})(\d_j\vec{n}\cdot\vec{\sigma})(\d_j\vec{n}\cdot\vec{\sigma})]\;,
\ee
which is related to the Hopf index.

\subsection{Bateman's construction}

We can introduce a complex vector (it would actually be $\vec{F}=\vec{E}+ic\vec{B}$, but we put $c=1$)
\be
\vec{F}\equiv \vec{E}+i\vec{B}.
\ee

Then Maxwell's equations become
\be
\vec{\nabla}\times \vec{F}=i\frac{\d}{\d t}\vec{F};\;\;\; \vec{\nabla}\cdot \vec{F}=0.
\ee

We can introduce an ansatz, that automatically satisfies $\vec{\nabla}\cdot \vec{F}=0$, namely 
\be
\vec{F}=\vec{\nabla}\a \times \vec{\nabla}\b.
\ee

The remaining equations of motion become 
\be\label{Batequ}
i\vec{\nabla}\times (\d_t \a \vec{\nabla}\b-\d_t\b\vec{\nabla}\a)=\vec{\nabla}\times \vec{F}\;,
\ee
solved by 
\be\label{Batequsol}
i (\d_t \a \vec{\nabla}\b-\d_t\b\vec{\nabla}\a)=\vec{F}=\vec{\nabla}\a\times \vec{\nabla}\b.
\ee
This implies that $\vec{F}^2=0$, so 
\bea
&&\vec{F}^2=i (\d_t \a \vec{\nabla}\b-\d_t\b\vec{\nabla}\a)(\vec{\nabla}\a\times \vec{\nabla}\b)=0\Rightarrow\cr
&&\vec{E}^2-\vec{B}^2=0,\;\;\;\; \vec{E}\cdot\vec{B}=0.
\eea
Note that then electromagnetic duality is simply $\a\rightarrow i\a$ or $\b\rightarrow i\b$. Also note that 
then for these solutions, the helicities $H_{ee}, H_{mm}, H_{em}$ and $H_{me}$ are all conserved. 

{\bf The Hopfion solution for electromagnetism}

The Hopfion solution for electromagnetism, in Bateman's construction, is defined by 
\bea
\a&=&\frac{A-1+iz}{A+it};\cr
\b&=&\frac{x-iy}{A+it}\cr
A&=& \frac{1}{2}(x^2+y^2+z^2-t^2+1).
\eea

In terms of two complex maps  $\phi,\theta:S^3\rightarrow S^2$, evolved in time from  the form (\ref{phiH}), 
we have the solution
\bea
\phi&=& \frac{Az+t(A-1)+i(tx-Ay)}{Ax+ty+i(A(A-1)-tz)}\cr
\theta&=&\frac{Ax+ty+i(Az+t(A-1))}{tx-Ay+i(A(A-1)-tz)}.
\eea
It is obtained by writing an ansatz for the relativistic field strength $F_{\mu\nu}$ and is dual, 
\be
\tilde F_{\mu\nu}=\frac{1}{2}\epsilon_{\mu\nu\rho\sigma}F^{\rho\sigma}\;,
\ee
that generalizes the  2-form on $S^3$ (\ref{2form}) to 4 dimensions, i.e. replacing the $i,j$ indices with $\mu,\nu$ indices. 
This turns out to solve Maxwell's equations, as we will 
shortly see (basically, since Mawell's equtions in vacuum become just $dF=0, d\tilde F=0$, or $F=dA$, $\tilde F=dC$).

Then the two maps reduce to a Hopf map with winding one (\ref{phiH}) at $t=0$,
\bea
\phi(t=0,x,y,x)&=& \frac{2(z-iy)}{2x+i(r^2-1)}=\phi_H(z,-y,x)\cr
\theta(t=0,x,y,z)&=&\frac{2(x+iz)}{-2y +i(r^2-1)}=\phi_H(x,z,-y).
\eea

Since the solution satisfies $\vec{F}^2=0$ by Bateman's construction, i.e. $\vec{E}\cdot \vec{B}=0$ and $\vec{E}^2-\vec{B}^2=0$, 
all its 4 helicities will be conserved. However, one finds that for it, 
\be
H_{ee}=H_{mm}\neq 0;\;\;\; H_{em}=H_{me}=0. 
\ee

\subsection{Ranada's construction}

This is an alternative construction of solutions, that turns out to admit a more general class of solutions, namely ones with $\vec{E}\cdot \vec{B}=0$, 
but $\vec{E}^2-\vec{B}^2$ arbitrary, so only with $H_{ee}$ and $H_{mm}$ always conserved. In terms of the ansatz, it is written by replacing the 
complex functions $\a$ and $\b$ with real functions. For more details of the construction, see \cite{arrayas2016knots}.

We can consider an ansatz for the electromagnetic fields in the vacuum,
\bea
\vec{B}(\vec{r},t)&=& g(\phi,\bar\phi) \vec{\nabla}\phi\times \vec{\nabla}\bar\phi\cr
\vec{E}(\vec{r},t)&=& g(\phi,\bar\phi) (\d_0\bar\phi\vec{\nabla}\phi-\d_0\phi\vec{\nabla}\bar\phi)\Rightarrow\cr
F_{\mu\nu}&=& g(\phi,\bar\phi)(\d_\mu\bar\phi\d_\nu\phi-\d_\nu\bar\phi\d_\mu\phi)\;,\label{ranada}
\eea
and the ansatz for the electric-magnetic dual fields, coming replacing $F_{\mu\nu}$ with $\tilde F_{\mu\nu}$, or $\vec{E}$ with $\vec{B}$,
\bea
\vec{E}(\vec{r},t)&=& g'(\theta,\bar\theta) \vec{\nabla}\theta\times \vec{\nabla}\bar\theta\cr
\vec{B}(\vec{r},t)&=& g'(\theta,\bar\theta) (\d_0\bar\theta\vec{\nabla}\theta-\d_0\theta\vec{\nabla}\bar\theta)\Rightarrow\cr
\tilde F_{\mu\nu}&=& g'(\theta,\bar\theta)(\d_\mu\bar\theta\d_\nu\theta-\d_\nu\bar\theta\d_\mu\theta).
\eea
We can choose the functions $g$ and $g'$ to be 
\be
g(\phi,\bar\phi)=\frac{\sqrt{a}}{2\pi i}\frac{1}{(1+\bar\phi\phi)^2};\;\;\;
g'(\theta,\bar\theta)=\frac{\sqrt{a}}{2\pi i}\frac{1}{(1+\bar\theta\theta)^2}.
\ee

We can immediately check that then, by construction, $\vec{E}\cdot\vec{B}=0$, but $\vec{E}^2-\vec{B}^2$ is not necessarily zero. We can also 
check that the ansatz gives
\be
\d_{[\mu}F_{\nu\rho]}=2\d_{[\mu}g \d_{\nu}\bar\phi\d_{\rho]}\phi=0\;,
\ee
and $\d^\mu F_{\mu\nu}=0$ is a further constraint, coming from electric-magnetic duality. 

A careful analysis shows that, if $\vec{E}\cdot \vec{B}=0$, the 2-form $F$ (and its dual) are {\em locally} decomposable as (Clebsh representation)
\be
F=dq\wedge dp\;,
\ee
where $p$ and $q$ are {\em real} functions. 

Note that, if $p$ and $q$ are {\em single-valued and well-defined in the whole of space}, then the magnetic 
helicity $\int \vec{A}\cdot\vec{B}$ is zero. 

Indeed, the decomposition $F=dq\wedge dp$ means that 
\be
\vec{B}=\vec{\nabla}p\times \vec{\nabla}q.
\ee
For nonzero magnetic helicity, i.e. when $p$ and $q$ are {\em not} well defined over the whole space, i.e. there are some singularities, then we can 
write 
\be
\vec{A}=p\vec{\nabla}q+\vec{\nabla}\chi\;,\label{vecAdecomp}
\ee
where $\chi$ is needed in order to absorb the singularities. Then 
\be
H_{mm}=\frac{1}{2\mu_0}\int \vec{A}\cdot \vec{B}=\frac{1}{2\mu_0}\int d^3x \vec{\nabla}\chi\cdot(\vec{\nabla} p\times \vec{\nabla}q).
\ee
If $p$ and $q$ are well defined over the whole space, there is no need to add $\vec{\nabla}\chi$, so this is the same as considering $\chi=0$ in the 
$H_{mm}$ above, hence we get a zero result.

In the representation $\vec{B}=\vec{\nabla}p\times \vec{\nabla}q$ and the ansatz for its dual, 
\be
\vec{E}=\vec{\nabla}u\times \vec{\nabla}v\;, 
\ee
$p,q,u,v$ are also called Euler potentials. In terms of the Ranada form (\ref{ranada}), we have 
\be
p=\frac{1}{1+|\phi|^2};\;\;\; q=\frac{arg(\phi)}{2\pi}\;,
\ee
and a similar relation for the dual variables. 

{\bf Non-null solutions}

Besides the null ($\vec{F}^2=0$) Hopfion solution already presented, 
there are a set of solutions with $\vec{E}\cdot \vec{B}=0$, but $\vec{E}^2-\vec{B}^2\neq 0$ that can be defined by giving a set of complex 
scalars $\phi_0^{(n,m)},\theta_0^{(l,s)}$ (each depending on two integers) at time $t=0$, defining $\vec{B}(t=0), \vec{E}(t=0)$, and then from the 
Maxwell's equations $\d^\mu F_{\mu\nu}=0$ finding the solutions at time $t\neq 0$. The solutions at $t=0$ are given by 
\bea
\phi_0&=& \frac{(X+iY)^{(n)}}{(Z+i(R^2-1)/2)^{(m)}}\cr
\theta_0&=&\frac{(Y+iZ)^{(l)}}{(X+i(R^2-1)/2)^{(s)}}\;,
\eea
where the $(m)$ index means we leave the modulus intact, but we raise the phase to the $m$ power. Consider the solution with $l=s=0$, 
which has $\vec{E}(\vec{r},t=0)=0$, and 
\bea
\vec{B}(\vec{r},t=0)&=&\frac{\sqrt{a}}{2\pi i}\frac{\vec{\nabla}\phi_0\times \vec{\nabla}\bar\phi_0}{(1+\bar \phi_0\phi_0)^2}=\sqrt{a}
\vec{\nabla}\left(\frac{1}{1+|\phi_0|^2}\right)\times\vec{\nabla}\left(\frac{arg(\phi)}{2\pi}\right)\cr
&=&\frac{\sqrt{a}}{2\pi}\vec{\nabla}\left(\frac{4Z^2+(R^2-1)^2}{(R^2+1)^2}\right)\times\vec{\nabla}[n\; arg(X+iY)-m\; arg(Z+i(R^2-1)/2)].\cr
&&
\eea

We can write the ansatz for $\vec{A}$ as well, in the gauge $A_0=$const, $\vec{\nabla}\cdot\vec{A}=0$, as 
\be
\vec{A}(\vec{r},t=0)=\frac{\sqrt{a}}{2\pi}\frac{4Z^2+(R^2-1)^2}{(R^2+1)^2}\times\vec{\nabla}[n\; arg(X+iY)-m\; arg(Z+i(R^2-1)/2)]
+\vec{\nabla}\chi\;,
\ee
where $\chi$ needs to cancel the singularities in the first term, and must satisfy (from $\vec{\nabla}\cdot\vec{A}=0$)
\be
\Delta\chi=-\vec{\nabla}\cdot\left[\frac{\sqrt{a}}{2\pi}\frac{4Z^2+(R^2-1)^2}{(R^2+1)^2}\times
\vec{\nabla}[n \;arg(X+iY)-m \;arg(Z+i(R^2-1)/2)\right].\label{chicond}
\ee

The time evolution of the solution is found from the equivalent of the Maxwell's equations, which in this formulation is the condition that 
$\vec{E}$ and $\vec{B}$ have the expressions in terms of $\phi_0$ and $\theta_0$ in {\em both} dual formulations, so that 
\be
\frac{\vec{\nabla}\phi_0\times\vec{\nabla}\bar\phi_0}{(1+\bar\phi_0\phi_0)^2}=\frac{1}{(1+\bar\theta_0\theta_0)^2}\left(\frac{\d \bar
\theta_0}{\d t}\vec{\nabla}\theta-\frac{\d \theta}{\d t}\vec{\nabla}\bar\theta_0\right)
\ee
and the relation with $\phi_0$ and $\theta_0$ interchanged.

\subsection{Solutions in nonlinear theories of electromagnetism}

In this subsection we prove that not only are the knotted solutions like the Hopfion solutions of electromagnetism, they are also solutions of 
{\em any} nonlinear theories that reduce to electromagnetism at small fields, theories like the Born-Infeld theory, or theories obtained by integrating 
out {\em any} light fields interacting with electromagnetism, like the Euler-Heisenberg Lagrangean.\footnote{We became aware that some of the 
conclusions in this subsection (relating to nonlinear eletromagnetism theories) were reached in a somewhat different way by 
\cite{Goulart:2016orx}, but we believe that argument is incomplete.}

To set up the notation, we use the formalism of Born and Infeld \cite{Born:1934gh} for analyzing their case (see \cite{Nastase:2015ixa} for 
a generalization of this analysis). Defining the quantities
\bea
F&\equiv& \frac{F_{\mu\nu} F^{\mu\nu}}{2b^2}=\frac{1}{b^2}(\vec{B}^2-\vec{E}^2);\cr
G&\equiv&
\frac{1}{8b^2}\epsilon^{\mu\nu\rho\sigma}F_{\mu\nu}F_{\rho\sigma}= \frac{1}{b^2}\vec{E}\cdot \vec{B}\;,
\eea
where $b$ is a dimensionful constant (of mass dimension 2), the BI Lagrangean is 
\be
{\cal L}=-b^2[\sqrt{1+F-G^2}-1].
\ee
We can define the analogous quantities to the ones of electromagnetism in a medium,
\be
\vec{H}\equiv -\frac{\d {\cal L}}{\d \vec{B}},\;\;\;\;
\vec{D}\equiv +\frac{\d {\cal L}}{\d \vec{E}}\;,
\ee
in terms of which the Maxwell equations take the same form as for electromagnetism in a medium,
\bea
\vec{\nabla}\times \vec{E}+\d_0 \vec{B}=0, && \vec{\nabla}\cdot \vec{B}=0\cr
\vec{\nabla}\times\vec{H}-\d_0\vec{D}=0, && \vec{\nabla}\cdot \vec{D}=0.
\eea
In the case of the BI theory, we obtain 
\be
\vec{H}=\frac{\vec{B}-G\vec{E}}{\sqrt{1-F+G^2}};\;\;\; 
\vec{D}=\frac{\vec{E}+G\vec{B}}{\sqrt{1-F+G^2}}\;,
\ee
so we explicitly see that for $F=G=0$, like is the case for the Hopfion and the knotted null solutions, we have $\vec{H}=\vec{B}$ and 
$\vec{D}=\vec{B}$, so the Maxwell equations reduce to the ones in vacuum. From the above, we can see that the same is true for any 
Lagrangean that is a function of $F$ and $G$ only, and contains electromagnetism as a limit of small fields, i.e. it can be written as an expansion
containing the Maxwell term plus higher order terms,
\be
{\cal L}=b^2\left[-\frac{F}{2}+\sum_{n\geq 2}\sum_{m\geq 0}c_{n,m}F^nG^m\right].\label{expan}
\ee
Indeed, then we have 
\be
\vec{H}=\vec{B}+{\cal O}(F,G);\;\;\;
\vec{D}=\vec{E}+{\cal O}(F,G)\;,
\ee
so if $F=G=0$ we obtain the usual Maxwell's equations in terms of $\vec{E}$ and $\vec{B}$.

An example of a nonlinear action is the Euler-Heisenberg action, obtained from integrating out the fermions in QED, which at one-loop 
gives (see \cite{Dunne:2004nc} for a review)
\bea
{\cal L}&=&\frac{\vec{E}^2-\vec{B}^2}{2}+\frac{2\a^4}{45m^4}[(\vec{E}^2-\vec{B}^2)^2+7(\vec{E}\cdot\vec{B})^2]\cr
&=&b^2\left[-\frac{F}{2}+\b b^2[F^2+7G^2]\right]\;,
\eea
where $b=m^2$ and $\b =2\a^2/45$. 

At higher loops, and when integrating out other fields, there are no explicit formulas for the Lagrangean 
written in terms of $F$ and $G$ only \cite{Dunne:2004nc}, however simply from Lorentz invariance, the Lagrangean {\em must} be written in terms of the 
only 2 Lorentz invariants made up of $F_{\mu\nu}$, namely $F$ and $G$. If we are at large distances (low energies), 
there are no covariant derivatives that can 
act on them. Moreover, since the nonlinear terms are quantum corrections, 
and the classical part is still Maxwell theory, the large distance Lagrangean must be of the form (\ref{expan}). 

Note that it is not surprising that {\em at the classical level} we have the null knottted solutions as solutions to Maxwell electromagnetism coupled to other
fields, since Maxwell theory is a consistent truncation of the full theory (meaning that there are only couplings quadratic in other fields, so the 
equations of motion of Maxwell theory solve the full theory). What is surprising is that {\em when taking into account the full quantum effects}, i.e. 
loops of the other fields, the knotted solutions remain solutions (which is not true for non-knotted solutions, for which generically $F,G\neq 0$ 
on-shell). 

So integrating out {\em any} fields (fermions, but also scalars, other gauge fields, etc.) coupled to electromagnetism, or simply having generic 
higher order corrections to it, the null knotted solutions are still solutions. Moreover, in the case of the BI action, it also 
arises in string theory, on the worldvolume of D3-branes, nonperturbative objects in string theory, by integrating out {\em string modes}, and 
then $b=1/\a'$ ($\a'=l_s^2$ is the string scale).

\section{Knotted solutions in fluid dynamics}

There has been a large literature on knots in fluids, starting with the 19th century. In fact, the theory of knots started this way. However, in 
terms of explicit knotted solutions of Euler's fluid equations (not to say of the viscous case, the Navier-Stokes equations), much less is known. 
This is due in part to the fact that, unlike Maxwell's equations, the Euler equations are already nonlinear (and going to the Navier-Stokes equations 
involves a further nonlinearity). In this section, we will review the equations of motion, invariants, and solutions that are known in fluid dynamics.

\subsection{Fluid equations and helicity}

A fluid is described by a density $\rho$, pressure $P$, velocity $\vec{v}$ and entropy density (per unit mass) $s$ fields. One can introduce also an 
outside potential (like for instance a gravitational potential) per unit mass $\pi$. 

The fluid equations are then:

-The Euler equation (with potential per unit mass $\pi$)
\be
\d_t\vec{v}+(\vec{v}\cdot\vec{\nabla})\vec{v}=-\frac{1}{\rho}\vec{\nabla}p\;(-\vec{\nabla}\pi)\;,\label{origeuler}
\ee

-The continuity equation
\be
\d_t\rho+\vec{\nabla}\cdot (\rho \vec{v})=0\;,
\ee

-The entropy equation (adiabaticity, or conservation of the entropy density per unit mass $s$),
\be
\d_t s+\vec{v}\cdot\vec{\nabla}s=0.
\ee

We will consider isentropic (constant entropy density) flow. Then $\d_t s=0$, which implies $\vec{v}\cdot \vec{\nabla}s=0$ from the entropy equation, 
so $s=s_0$ everywhere and at any time. That means that we can deal with only the Euler equation and the continuity equation. 

Taking $\vec{\nabla}\times $ of the Euler equation, we obtain the vorticity equation
\be
\d_t\vec{\omega}+\vec{\nabla}\times (\vec{\omega}\times \vec{v})=0.
\ee

Since the enthalpy $H=U+PV$ and $dU=TdS-PdV+d\Pi$, it follows that $dh=dp/\rho+tds+d\pi$, where $h=H/M$ is enthalpy per unit mass. 
If $ds=0$, then we have 
\be
\vec{\nabla}h=\frac{1}{\rho}\vec{\nabla}p+\vec{\nabla}\pi. 
\ee

More generally, we have 
\be
\delta h =\frac{\delta p}{\rho}+\delta \pi=c_s^2\frac{\delta \rho}{\rho}+\delta \pi\;,
\ee
where $c_s^2=\d p/\d \rho$ is the sound speed squared. 

Then, the Euler equation becomes 
\be
\d_t \vec{v}+(\vec{v}\cdot\vec{\nabla})\vec{v}+\vec{\nabla}h=0.\label{euler}
\ee

And, since now similarly (for $\d_t\pi=0$)
\be
\d_t \rho =\frac{1}{c_s^2}\d_t h\;,
\ee
and (if $\vec{\nabla}\pi=0$)
\be
\vec{\nabla}\rho=\frac{1}{c_s^2}\vec{\nabla}h\;,
\ee
the continuity equation becomes
\be
\d_t h+\vec{v}\cdot\vec{\nabla}h+c_s^2\vec{\nabla}\cdot\vec{v}=0.\label{cont}
\ee

The fluid dynamics equations are then (\ref{euler}) and (\ref{cont}).

Moreover, using the relation (valid for any vector)
\be
(\vec{v}\cdot\vec{\nabla})\vec{v}=-\vec{v}\times(\vec{\nabla}\times\vec{v})+\vec{\nabla}(\vec{v}^2/2)\;,
\ee
and the fact that $\vec{\omega}=\vec{\nabla}\times \vec{v}$,
the Euler equation is written still as 
\be
\d_t\vec{v}+\vec{\omega}\times\vec{v}+\vec{\nabla}\left(\frac{\vec{v}^2}{2}+h\right)=0.
\ee

{\bf Conserved helicity}

Define the quantity
\be
H=\int d^3x \vec{v}\cdot \vec{\omega}=\int d^3x \vec{v}\cdot(\vec{\nabla}\times\vec{v})=\int d^3x \epsilon^{ijk} v_i\d_j v_k\;,\label{fluidhel}
\ee
which is analogous to the magnetic helicity, being also written like a Chern-Simons term, this time for $v_i$ instead of $A_i$. However, now $v_i$ 
is an observable, so there is no equivalent of the gauge invariance in the electromagnetic case. But like for $H_{mm}$, $v^i=v^i(\vec{x},t)$ is a 
function of spacetime, not only of the space that we integrate over. It means that it is again not clear if $H_{mm}$ is conserved (independent of time), 
and it doesn't necessarily have integer values. 

From the original Euler's equation (\ref{origeuler}), we can find the equation
\be
\d_t(\vec{v}\cdot\vec{\omega})-\vec{\nabla}\cdot\left[\vec{\omega}\left(\frac{\vec{v}^2}{2}-\int \frac{dp}{\rho}-\pi\right)-\vec{v} (\vec{v}\cdot\vec{\omega})\right]=0.
\ee
But that says that the time derivative of the integrand of $H$ is a total derivative. That means that $\d_t H$ is a boundary term, which can be made to 
vanish, for a localized flow, hence $H$ is actually conserved in time. Knotted solutions, i.e. solutions for which the velocity field $v^i(\vec{x},t)$ 
at a given time will have a nonzero topological index (linking number, or knottedness), will have a nonzero $H$. 

Except for the issue if gauge invariance, we see an important analogy, as far as knotted solutions go, between fluid dynamics and 
electromagnetism:  $\vec{v}$ is the analog of $\vec{A}$, so $\vec{B}$ is the analog of the vorticity $\vec{\omega}$, and $H$ is 
the analog of $H_{mm}$.

We can guess from the electromagnetic analogy that we can construct solutions with $H\neq 0$, that will be knotted, specifically having 
nonzero Hopf index. 

\subsection{Solutions with Hopf index in fluid dynamics}

In \cite{kuznetsov1980topological}, the equations of motion of the incompressible fluid, i.e. the fluid satisfying 
\be
\vec{\nabla}\cdot \vec{v}=0\;,
\ee
were analyzed.

In this case, it was found that the fluid flow is Hamiltonian, in the sense that there
are parameters for which the equations of motion are first order in time. 

The construction of  \cite{kuznetsov1980topological} follows the same logic as the Ranada construction in electromagnetism. 

We define the {\em Clebsch variables} $\lambda$ and $\mu$ by the equations
\bea
\frac{\d\lambda}{\d t}&=&-(\vec{v}\cdot \vec{\nabla})\lambda\cr
\frac{\d\mu}{\d t}&=&-(\vec{v}\cdot\vec{\nabla})\mu\;,\label{lambdamu}
\eea
such that the velocity field is 
\be
\vec{v}=\lambda\vec{\nabla}\mu +\vec{\nabla}\Phi\;,\label{velclebsch}
\ee
where $\Phi$ is the fluid potential (for a vorticity-free field, $\omega=\vec{\nabla}\times \vec{v}=0$, we would have only the $\Phi$ term). 

The relation (\ref{velclebsch}) is the same as the ansatz for the electromagnetic $\vec{A}$ in (\ref{vecAdecomp}). Like in that case, we have 
that if $\lambda$ and $\mu$ are single valued, i.e. if they are {\em real} functions (there are Riemann sheets only on the 
complex plane) with no singularities, there are no knots, and in fact  the fluid helicity $H$ (\ref{fluidhel}) is zero, just like $H_{mm}$ was zero.

Then, from the incompressibility condition $\vec{\nabla}\cdot \vec{v}=0$, we obtain that $\Phi$ must satisfy
\be
\Delta\Phi=-\vec{\nabla}\cdot(\lambda\vec{\nabla}\mu).
\ee

Note that the ansatz (\ref{velclebsch}) contains no time derivatives, so can be taken at each moment in time. The time evolution is given by 
the Hamiltonian relations (\ref{lambdamu}), or by the Euler equation for the velocity, and is thus highly nonlinear. 

Calculating $\d \vec{v}/\d t$ by substituting the equations of motion (\ref{lambdamu}) for $\lambda,\mu$ and then re-constituting $\vec{v}$, we obtain
\bea
&&\frac{\d \vec{v}}{\d t}=-(\vec{v}\cdot \vec{\nabla})(\lambda\vec{\nabla}\mu)+\vec{\nabla}\frac{\d \phi}{\d t}\Rightarrow\cr
&& \frac{\d \vec{v}}{\d t}+(\vec{v}\cdot\vec{\nabla})\vec{v}=\left(\vec{v}\cdot\vec{\nabla}+\frac{\d}{\d t}\right)\vec{\nabla}\phi.
\eea
Taking the $\vec{\nabla}\times$ of this equation, one gets the 
vorticity equation,
\be
\d_t\vec{\omega}=\vec{\nabla}\times(\vec{v}\times \vec{\omega})\;,
\ee
as one should.

In these Clebsch variables, the ansatz for the vorticity is
\be
\omega=\vec{\nabla}\times \vec{v}=\vec{\nabla}\lambda \times \vec{\nabla}\mu\;,
\ee
which maps to the analog ansatz for the magnetic field in the Ranada construction,
\be
\vec{B}=\vec{\nabla}\lambda\times \vec{\nabla}\mu.
\ee

{\bf Solutions with Hopf index}

Since we work with a vorticity $\omega$, which by 
construction  satisfies $\vec{\nabla}\cdot\vec{\omega}=0$, since it is the curl of something, we can write it as 
\be
\omega^\a=\epsilon_{ijk}\epsilon^{\a\b\gamma} n^i\d_\b n^j \d_\gamma n^k.
\ee
Indeed, since $n^i n^i=1$, it follows that $\vec{\nabla}\cdot\vec{\omega}\equiv \d_\a \omega^\a=0$. Then moreover, consider {\em only 
$\vec{n}$'s that go to a constant at infinity} (see the discussion about the Hopf index), so that the map $\vec{n}$ is 
compactified to a map $\vec{n}:S^3\rightarrow S^2$, characterized by the Hopf index. In fact, now the helicity $H$ 
is actually proportional to the Hopf index, and in the superfluid case the proportionality is \cite{kuznetsov1980topological}
\be
H=64\pi^2\left(\frac{\hbar}{4m}\right)^2 N\;,
\ee
where $N$ is the Hopf index. Since we saw that for localized flows $H$ is invariant in time, if we give an initial condition with $N=1$ ($H\neq 0$). 
it will stay so in time. But since $\vec{\omega}$ is written in terms of $\vec{n}$, all one needs is to give an example with Hopf index one 
in terms of $\vec{n}$. 

For instance, \cite{kuznetsov1980topological} chooses ($\sigma^i$ are Pauli matrices)
\bea
\vec{n}\cdot\vec{\sigma}&=&q^\dagger\sigma_3q\cr
q&=&(1-i\vec{r}\cdot \vec{\sigma})(1+i\vec{r}\cdot \vec{\sigma})^{-1}.\label{choice}
\eea

But this parametrization by $\vec{n}$ is simply related to the Clebsch variables, which are just the angles on $S^2$, $\theta,\phi$. 
Thus using such a parametrization 
\be
(n_1,n_2,n_3)=(\sin \theta\sin\phi,\sin\theta\cos\phi,\cos\theta)\;,
\ee
we find 
\be
\vec{\omega}=\vec{\nabla}\cos\theta\times\vec{\nabla}\phi\Rightarrow \omega^\a=\epsilon^{\a\b\gamma}\d_\b \cos\theta\d_\gamma\phi\;,
\ee
so $\lambda=\cos\theta,\mu=\phi$.

\section{Euler fluid as restricted electromagnetism}

We saw that $\vec{B}=\vec{\nabla}\times \vec{A}$ in electromagnetism is analogous to $\vec{\omega}=\vec{\nabla}\times \vec{v}$ in a fluid, 
in terms of constructions that have nonzero Hopf index and give a nonzero helicity. An extension of this analogy was intensively  used 
(see, for instance, \cite{Kholodenko:2014apa,Kholodenko:2014wfa,Boozer}), 
from magneto-hydrodynamics, i.e. a fluid coupled to electromagnetism, to hydrodynamics only. That is, consider $\vec{v}$ mapped to
$\vec{v}$, but then $\vec{B}$ mapped to $\vec{\omega}$. Then the hydrodynamics configurations are restricted magnetohydrodynamics configurations, 
ones where we remember that $\vec{B}=\vec{\omega}=\vec{\nabla}\times \vec{v}$. 

But one can construct another map, simply between the Euler fluid and electromagnetism. Part of it is clearly the identification of $\vec{v}$ with 
$\vec{A}$, so $\vec{\omega}$ with $\vec{B}$. In fact, we saw that in this way, both the helicities map to each other ($H$ to $H_{mm}$), 
and the constructions of singular solutions. But we can extend this map to a full map, defining electromagnetic variables from the fluid variables,
\bea
\vec{B}&=&\vec{\omega}=\vec{\nabla}\times\vec{v}\cr
\vec{E}&=&-\d_t \vec{v}-\vec{\nabla}h\;,
\eea
which means that really 
\be
\vec{A}=\vec{v};\;\;\; \phi=h.
\ee
However, there is no gauge invariance now, since $\vec{v}$ is a physical (measurable) quantity, and unlike $\phi$,  $h$ cannot be put to zero
in the absence of sources, but rather only to a constant $h_0$, which is enough.   
Thus  the analogy is not complete, but nevertheless  it can be used formally.

One thing which works is that we obtain part of the Maxwell equations. 
The zero divergence of the magnetic field, 
\be
\vec{\nabla}\cdot \vec{B}=0\;,
\ee
is as usual, simply the result of writing $\vec{B}$ as the $\vec{\nabla}\times $ of something. Also from the definition, we have 
\be
\vec{\nabla}\times \vec{E}=-\frac{\d}{\d t} \vec{B}\;,
\ee
since both of these equations are really the Bianchi identities. Expressing it in differential forms,  $dF=0$, which follows from $F=dA$, and
is  valid since we have identified  $A_\mu=(h,\vec{v})$.

But the Euler equation, the time evolution of the fluid velocity, is a nonlinear equation, and cannot be rewritten as a combination of the 
linear Maxwell's equations of motion $d*F=0$. So the time evolution does not quite work, unless one is willing to linearize the equation, and then, 
using Euler's equation (\ref{euler}), we find 
\be
\vec{E}=(\vec{v}\cdot\vec{\nabla})\vec{v}=\vec{\omega}\times \vec{v}+\vec{\nabla}\left(\frac{\vec{v}^2}{2}\right)\;,
\ee
which could be further expressed in terms of the magnetic variables as 
\be
\vec{E}=\vec{B}\times \vec{A}+\vec{\nabla}\left(\frac{\vec{A}^2}{2}\right)\;,\label{nonlinear}
\ee
and at the linear level this simply implies that $\vec{E}=0$. Thus linearized fluid dynamics becomes pure magnetism under this map. 

It is interesting to note that the continuity equation (\ref{cont}), i.e.
\be
\d_t h+(\vec{v}\cdot\vec{\nabla})h+c_s^2\vec{\nabla}\cdot\vec{v}=0\;,
\ee
becomes in electromagnetism language
\be
\d_t \phi+(\vec{A}\cdot \vec{\nabla})\phi+c_s^2\vec{\nabla}\cdot \vec{A}=0.\label{lorentz}
\ee
As we see, linearizing it, we obtain the {\em Lorenz gauge condition}.

One could solve the Lorenz gauge condition like a radiation gauge, by
\be
\phi=h={\rm const.};\;\;\; \vec{\nabla}\cdot \vec{A}=\vec{\nabla}\cdot \vec{v}=0\;,
\ee
and this would correspond in the fluid to an incompressible ideal fluid.

\section{New solutions in electromagnetism}

We can use the map from fluid dynamics to electromagnetism to find additional  electromagnetic solutions. 
The first observation is that, since in any case for the time evolution we have on one side the Euler equation, and on the other 
the $d*F=0$ equation, we can forget about the continuity equation (which can be thought of giving  time evolution to  $h$, the enthalpy density), 
hence forget about the radiation (or Lorenz) gauge condition in electromagnetism. This is in accordance with the fact that we are not in Lorenz gauge in the 
case of the Hopfion type solutions. 

To consider more general solutions suggested by the fluid map, we simply give an initial condition (at $t=0$) via the map, and then evolve in time 
with $d*F=0$. We can then solve $\vec{E}$ and $\vec{B}$ in terms of both $\phi$ and $\theta$. 
Therefore we write at $t=0$ (thus automatically satisfying $\vec{\nabla}\cdot \vec{B}=0$)
\be
B^a=\epsilon_{ijk}\epsilon^{abc}n^i \d_b n^j \d_c n^k\;,
\ee
with the same $n^i$ as \cite{kuznetsov1980topological}. We need to also put a similar ansatz for $\vec{E}$ at $t=0$, i.e.
\be
E^a=\epsilon_{ijk}\epsilon^{abc} \tilde n^i\d_b \tilde n^j\d_c\tilde n^k\;,
\ee
which automatically satisfies $\vec{\nabla}\cdot \vec{E}=0$. 

We can take the above as just initial conditions (for some appropriate vectors $n^i,\tilde n^i$, for instance the one in (\ref{choice})) 
for the Maxwell's equations. Or, more completely, we assume that the above forms for $B^a$ and $E^a$ are valid at all times (since then 
we automatically satisfy $\vec{\nabla}\cdot\vec{E}=\vec{\nabla}\cdot \vec{B}=0$), and we have $\vec{n}=\vec{n}(t)$ and $\vec{\tilde n}=\vec{\tilde n}(t)$. 

Then the Bianchi identity $\vec{\nabla}\times \vec{E}=-\d_t \vec{B}$ becomes
\be
\epsilon_{ijk}[2\d_e(\tilde n^i\d_a \tilde n^j\d_e\tilde n^k)+\epsilon^{abc}\d_t(n^i\d_b n^j\d_c n^k)]=0
\ee
and the equation of motion $\vec{\nabla}\times \vec{B}=\d_t \vec{E}$ becomes
\be
\epsilon_{ijk}[2\d_e(n^i\d_a n^j\d_e n^k)-\epsilon^{abc}\d_t(\tilde n^i\d_b\tilde n^j\d_c\tilde n^k)]=0.
\ee
These are two equations relating $n^i(t),\tilde n^i(t)$, which however must be solved numerically. Note that in the fluid case, the single vector $\vec{n}$ 
satisfies an equation derived from the Euler equation \cite{kuznetsov1980topological}, 
\be
\d_t\vec{n}+(\vec{v}\cdot\vec{\nabla})\vec{n}=0\;,
\ee
where $\vec{v}$ is defined by $\vec{\omega}$, thus by $\vec{n}$ itself,
but now we have two nonlinear equations for the two vectors.

\subsection{Solutions by conformal transformations}

In \cite{arrayas2016knots}, it is shown that if one takes a  special conformal transformation, specifically an inversion,
\be
x^\mu\rightarrow a \frac{x^\mu}{x^2}
\ee
on the constant RS field
\be
\vec{F}=\vec{E}+i\vec{B}=\vec{e}_x+i\vec{e}_y\;,
\ee
followed by an imaginary time translation, $t\rightarrow t-i$, we obtain the Hopfion solution with $\vec{F}^2=0$, i.e. 
$\vec{E}^2=\vec{B}^2$ (since this is true also before the transformations). The transformation law on $\vec{F}$ is 
\be
\vec{F}'(x)=\frac{1}{(t^2-r^2)^3}[-(t^2-r^2)\tilde{\vec{F}}(\tilde x(x))+2\vec{r}\times(\vec{r}\times\tilde{\vec{F}}(\tilde x(x)))-2it\vec{r}\times\tilde{\vec{F}}
(\tilde x(x))]\;,
\ee
where $\tilde{\vec{F}}=\vec{E}-i\vec{B}$. The final solution is written as 
\be
\vec{F}=\frac{1}{[(t-i)^2-r^2]^3}\begin{pmatrix}(x-iy)^2-(t-i-z)^2\\i(x-iy)^2+i(t-i-z)^2\\-2(x-iy)(t-i-z)\end{pmatrix}.
\ee

But we can consider similar transformations on an RS field that has only $\vec{E}\cdot\vec{B}=0$, but $\vec{E}^2\neq \vec{B}^2$, 
i.e. for which $\vec{F}^2$ is purely real. For instance, start with 
\be
\vec{F}=i\vec{e}_y\;,
\ee
and apply the same conformal transformation, to obtain 
\be
\vec{F}=\frac{1}{[t^2-r^2]^3}\begin{pmatrix} 2tz -2ixy\\ it^2+i(x^2-y^2+z^2)\\ -2tx -2i yz\end{pmatrix}.
\ee
Then $\vec{F}^2=[t^2-r^2]^{-4}$ is purely real, so still $\vec{E}\cdot\vec{B}=0$. There is a singular point at $x=z=0$, $t=y$ (moving at the 
speed of light). Doing a shift $t\rightarrow t-i$, we obtain 
\be
\vec{F}=\frac{1}{[(t-i)^2-r^2]^3}\begin{pmatrix} 2(t-i)z -2ixy\\ i(t-i)^2+i(x^2-y^2+z^2)\\ -2(t-i)x -2i yz\end{pmatrix}.
\ee
Now however, $\vec{F}^2=[(t-i)^2-r^2]^{-4}$ is not real anymore, so $\vec{E}\cdot \vec{B}\neq 0$. Nevertheless, this is now another 
solution of Maxwell's equations, since we performed the same isometries of the equations of motion as before.

\section{(Free) electromagnetism as a a fluid}

The fluid description is a large distance expansion for any field theory. The expansion is an expansion of the energy-momentum tensor in 
derivatives of the velocity (or 4-velocity $u^\mu$, in a relativistic formulation). The first order term is the perfect fluid term, with no derivatives on 
the velocity, and it leads to the Euler equation. If we consider the terms with $\d_\mu u_\nu$ in $T_{\mu\nu}$, we obtain the (relativistic) 
Navier-Stokes equation, etc. 

This program can be applied in principle to any quantum system \cite{Nastase:2015ljb,Endlich:2010hf,Dubovsky:2011sj,Berezhiani:2016dne}. 
Under certain conditions (like no vorticity of the fluid), a quantum system can be effectively described by a scalar field theory, which then can be 
used to define a fluid energy-momentum tensor \cite{Endlich:2010hf,Dubovsky:2011sj,Berezhiani:2016dne}. Or perhaps the system is 
already described well by a scalar field theory like DBI, and then one can do the same \cite{Nastase:2015ljb}. In the case of a scalar 
theory, the 4-velocity of the effective fluid can be defined as 
\be
u^\mu=\frac{\d^\mu\phi}{\sqrt{-(\d_\mu\phi)^2}}\;,
\ee
defined such that $u^\mu u_\mu=-1$ as usual, and then one can identify the density $\rho$, pressure $P$, as well as higher order coefficients 
(in the derivative expansion) from the energy-momentum tensor $T_{\mu\nu}$. 
Note that one has to at least restrict to a class of classical solutions of the field theory for the map to work; in this case, we need to have 
$(\d_\mu\phi)^2<0$.

However, in the scalar field case, a conformal theory has 
a traceless Belinfante tensor (the usual energy-momentum tensor obtained from varying the metric), ${T^\mu}_\mu=0$, only in $d=2$, and 
to make ${T^\mu}_\mu=0$, one needs to add to it a total derivative term $T_{\mu\nu}\rightarrow T_{\mu\nu}+\d^\rho J_{\rho\mu\nu}$, so 
there are some subtleties in the identification of $\rho, P$, etc. \cite{Nastase:2015ljb}.

When writing the derivative expansion, of a perfect fluid, plus a first order viscous fluid term, 
\bea
T_{\mu\nu}&=&\rho u_\mu u_\nu +P(g_{\mu\nu}+u_\mu u_\nu)+\pi_{\mu\nu}\cr
\pi_{\mu\nu}&=&-2\eta\left[\frac{\nabla_\mu u_\nu+\nabla_\nu u_\mu}{2}+\frac{a_\mu u_\nu+a_\nu u_\mu}{2}
-\frac{1}{3}(\nabla^\rho u_\rho)(g_{\mu\nu}+u_\mu u_\nu)\right]\cr
&&-\zeta(\nabla^\mu u_\mu)
(g_{\mu\nu}+u_\mu u_\nu)+...\;,
\eea
where $a_\mu=u^\rho \nabla_\rho u_\mu$ (the expansion is in the Landau frame, $u^\mu \pi_{\mu\nu}=0$, and $\eta$ is the shear viscosity, 
$\zeta$ is the bulk viscosity), the equations of motion of the fluid, namely the relativistic version of the  Euler equation (extended to the Navier-Stokes
equation by the inclusion of $\pi_{\mu\nu}$) and the continuity equation, are all obtained from the conservation of the energy-momentum tensor,
\be
\d^\mu T_{\mu\nu}=0.
\ee
At zeroth order (ignoring the viscous term $\pi_{\mu\nu}$), one obtains the equation
\be
u_\mu u_\nu \d^\mu(\rho+P)+(\rho+P)(\d^\mu u_\mu) u_\nu +(\rho+P)u_\mu \d^\mu u_\nu +\d_\nu P=0.
\ee

In the non-relativistic limit, $u^\mu\simeq (1,\vec{v})$, $|\vec{v}|^2\ll 1$, $P\ll \rho$, the $\nu=0$ component gives the continuity equation,
\be
\d_t\rho+\vec{\nabla}\cdot(\rho\vec{v})=0\;,
\ee
whereas the $\nu=i$ component gives the equation
\be
\vec{v}\cdot \left(\d_t\rho+\vec{\nabla}\cdot(\rho\vec{v})\right)+\rho\frac{\d \vec{v}}{\d t}
+\rho(\vec{v}\cdot\vec{\nabla})\vec{v}+\vec{\nabla}P=0\;,
\ee
which after using the continuity equation reduces to the Euler equation (at zero outside potential). 

Note however that the nonrelativistic limit is not the only condition that results in the continuity equation and the Euler equations. 
It is in fact also possible to consider another, more unusual, type of fluid, a ``null pressureless fluid", or ``null dust", with $P=0$, but $u^\mu u_\mu=0$, 
specifically $u^\mu=(1,\vec{v})$, $\vec{v}^2=1$. Under this assumption, the same $\nu=0$ and $\nu=i$ equations are obtained, so the 
same continuity equation and Euler equation. 

We want then to describe free Maxwell electromagnetism as a fluid. In this case, the free Maxwell electromagnetism does have a traceless 
energy-momentum tensor, so the subtlety in the scalar case does not appear. 
As in the scalar case, we need some restriction on the set of classical solutions. 
We want to identify the Maxwell energy-momentum tensor with the fluid one, so we need to identify (we consider $u^\mu=(1,\vec{v})$)
\bea
T_{00}&=&\frac{1}{2}(\vec{E}^2+\vec{B})^2\equiv \rho\cr
T_{0i}&=&[\vec{E}\times\vec{B}]_i\equiv \rho v_i\cr
T_{ij}&=&-\left[E_iE_j+B_iB_j-\frac{1}{2}(\vec{E}^2+\vec{B}^2)\delta_{ij}\right]\equiv \rho v_i v_j.
\eea
Note that we have assumed already that $P=0$, since otherwise we can find no matching. 
From the first two relations, we find ($P=0$ and)
\be
\rho\leftrightarrow \frac{1}{2}(\vec{E}^2+\vec{B}^2);\;\;\;\;
v_i\leftrightarrow\frac{[\vec{E}\times \vec{B}]_i}{\frac{1}{2}(\vec{E}^2+\vec{B}^2)}.\label{velmap}
\ee
But then the third relation is not always satisfied. Only for particular configurations it is. But we always have 
$T_{ij}\delta^{ij}=T_{00}$ for electromagnetism (i.e., ${T^\mu}_\mu=0$, as we can check), 
which implies (if there is a $v_i$ that satisfies the third relation) that $\vec{v}^2=1$, so indeed the fluid is null. 

But for the velocity defined by (\ref{velmap}) we find that $\vec{v}^2=1$ implies
\be
(\vec{E}^2-\vec{B}^2)^2+4(\vec{E}\cdot\vec{B})^2=0\Rightarrow \vec{E}\cdot \vec{B}=0\;\;\; {\rm and}\;\;\; \vec{E}^2-\vec{B}^2=0\;,\label{conds}
\ee
so these conditions are necessary. We find that they are also sufficient for matching the energy-momentum tensor with the fluid one.

For instance, for constant perpendicular fields, $\vec E= E\hat x$ and  $\vec B=B\hat y$, with $E=B$, we find the trivial solution of the Euler equations
\be
\rho=E^2,\;\;\; \vec v=  \hat z\;\; \Rightarrow \;\;T_{0i} = E^2 \hat z, \qquad T_{ij}= (0,0, E^2).
\ee

\section{New solutions in fluid dynamics}

We can now derive new solutions of the fluid dynamics equations, namely the Euler and continuity equations.

\subsection{New solutions of null fluid type}

Using the map of electromagnetism to fluid dynamics from section 6, valid for configurations satisfying $\vec{F}^2=0$, we can 
map electromagnetic solutions to fluid dynamics. 

An important observation about this map is that, while the Maxwell equations were linear (free theory), the fluid Euler equations are nonlinear. 
The reason is that the fluid ansatz for the energy-momentum tensor is nonlinear in $\rho$, $v_i$. The implication is that, by mapping linear solutions
of electromagnetism to fluid dynamics, we obtain nontrivial solutions, that would have been difficult to guess otherwise. 

{\bf Hopfion solution}

The Hopfion solution satisfies the condition $\vec{F}^2=0$, so we can map it to fluid dynamics. 
For the Hopfion, the electric field vector is given by ($r^2=x^2+y^2+z^2$)
 
\bea
E_x&=& 
\frac{2 t^3 (t-x y-z)-\frac{3}{2} t^2
   \left(-t^2+r^2+1\right) \left(t^2-2 t
   z-2x^2+r^2-1\right)}{2 \left(\frac{1}{4}
   \left(-t^2+r^2+1\right)^2+t^2\right)^3}\cr
&&+\frac{   \frac{3}{2} t (-t+x y+z)
   \left(-t^2+r^2+1\right)^2}{2 \left(\frac{1}{4}
   \left(-t^2+r^2+1\right)^2+t^2\right)^3}\cr
   &&\frac{+\frac{1}{8}
   \left(-t^2+r^2+1\right)^3 \left(t^2-2 t
   z-2x^2+r^2-1\right)}{2 \left(\frac{1}{4}
   \left(-t^2+r^2+1\right)^2+t^2\right)^3},\cr
E_y&=& 
   \frac{3 t^2 (t+x y-z)
   \left(-t^2+r^2+1\right)-\frac{3}{4} t
   \left(-t^2+r^2+1\right)^2 \left(t^2-2 t
   z+r^2-2y^2-1\right)}{2 \left(\frac{1}{4}
   \left(-t^2+r^2+1\right)^2+t^2\right)^3}\cr
&&+\frac{   -\frac{1}{4} (t+x y-z)
   \left(-t^2+r^2+1\right)^3+t^3 \left(t^2-2 t
   z+r^2-2y^2-1\right)}{2 \left(\frac{1}{4}
   \left(-t^2+r^2+1\right)^2+t^2\right)^3},\cr
E_z&=&   
   \frac{t^3 (y
   (t-z)+x)+\frac{3}{2} t^2 (-t x+x z+y)
   \left(-t^2+r^2+1\right)}{\left(\frac{1}{4}
   \left(-t^2+r^2+1\right)^2+t^2\right)^3}\cr
   &&+\frac{-\frac{3}{4} t (y (t-z)+x)
   \left(-t^2+r^2+1\right)^2}{\left(\frac{1}{4}
   \left(-t^2+r^2+1\right)^2+t^2\right)^3}\cr
&&\frac{   +\frac{1}{8} (t x-x z-y)
   \left(-t^2+r^2+1\right)^3}{\left(\frac{1}{4}
   \left(-t^2+r^2+1\right)^2+t^2\right)^3}
\eea

\begin{figure}[ht!]
\centering
\includegraphics[width=200bp]{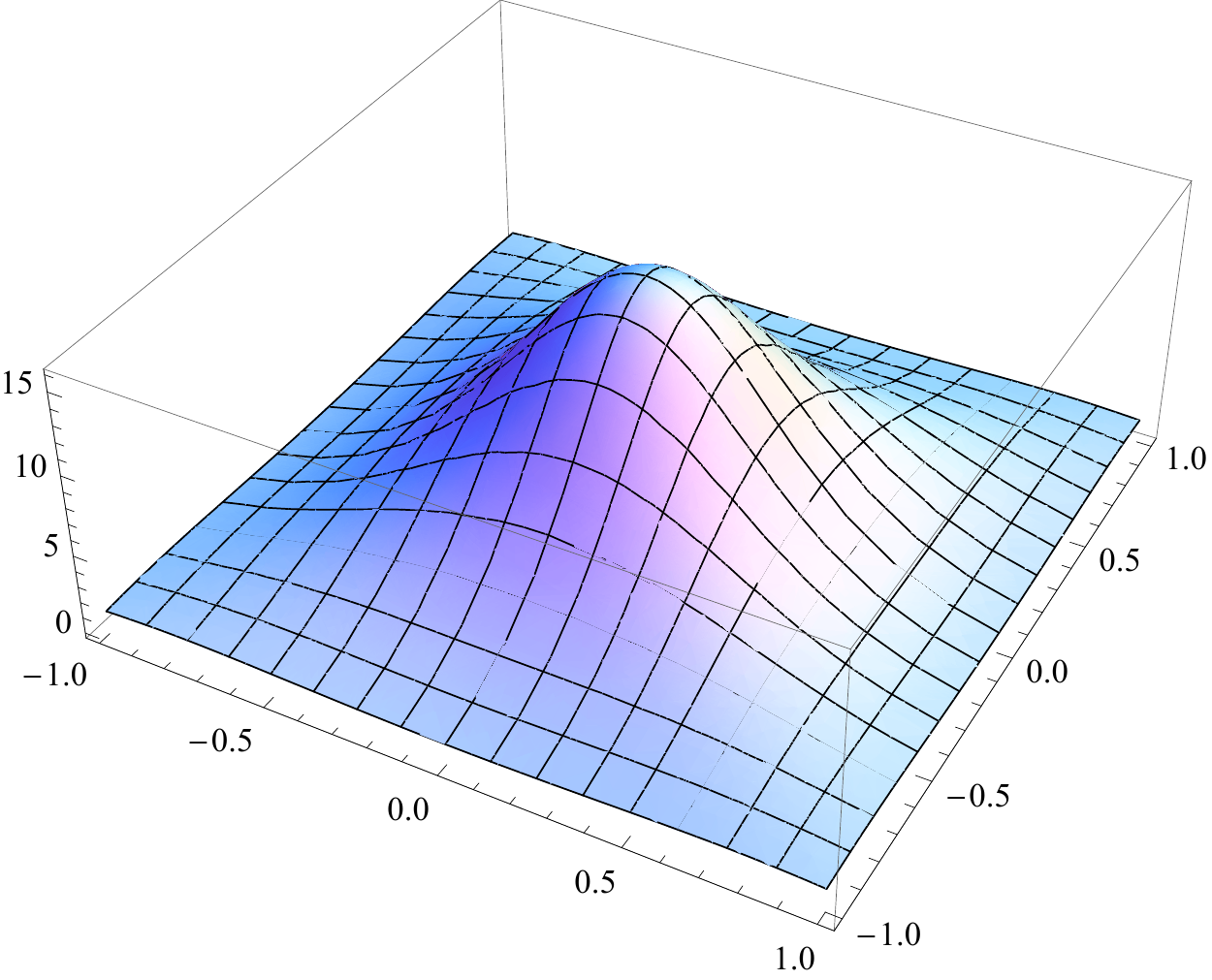}
\caption{The dependence of the density on the space-time coordinates.  The dependence on $t$ and $r=\sqrt{x^2+y^2}$ for z=0.}
\label{energydensity2}
\end{figure}

The magnetic field vector is given by 
\bea
B_x&=&
\frac{3 t^2 (t-x y-z) \left(-t^2+r^2+1\right)-\frac{3}{4} t
   \left(-t^2+r^2+1\right)^2 \left(t^2-2 t
   z-2x^2+r^2-1\right)}{2 \left(\frac{1}{4}
   \left(-t^2+r^2+1\right)^2+t^2\right)^3}\cr
&& +\frac{  +\frac{1}{4} (-t+x y+z)
   \left(-t^2+r^2+1\right)^3+t^3 \left(t^2-2 t
   z-2x^2+r^2-1\right)}{2 \left(\frac{1}{4}
   \left(-t^2+r^2+1\right)^2+t^2\right)^3},\cr
B_y&=&   
   \frac{-2 t^3 (t+x
   y-z)+\frac{3}{2} t^2 \left(-t^2+r^2+1\right) \left(t^2-2 t
   z+r^2-2y^2-1\right)}{2 \left(\frac{1}{4}
   \left(-t^2+r^2+1\right)^2+t^2\right)^3}\cr
   &&+\frac{+\frac{3}{2} t (t+x y-z)
   \left(-t^2+r^2+1\right)^2}{2 \left(\frac{1}{4}
   \left(-t^2+r^2+1\right)^2+t^2\right)^3}\cr
   &&\frac{-\frac{1}{8}
   \left(-t^2+r^2+1\right)^3 \left(t^2-2 t
   z+r^2-2y^2-1\right)}{2 \left(\frac{1}{4}
   \left(-t^2+r^2+1\right)^2+t^2\right)^3},\cr
B_z&=&   
   \frac{t^3 (t x-x  z-y)+\frac{3}{2} t^2 (y (t-z)+x)
   \left(-t^2+r^2+1\right)}{\left(\frac{1}{4}
   \left(-t^2+r^2+1\right)^2+t^2\right)^3}\cr
   &&+\frac{+\frac{3}{4} t (-t x+x z+y)
   \left(-t^2+r^2+1\right)^2}{\left(\frac{1}{4}
   \left(-t^2+r^2+1\right)^2+t^2\right)^3}\cr
   &&\frac{-\frac{1}{8} (y (t-z)+x)
   \left(-t^2+r^2+1\right)^3}{\left(\frac{1}{4}
   \left(-t^2+r^2+1\right)^2+t^2\right)^3}
\eea
We can check that indeed
\be
\vec E\cdot \vec B =0; \qquad  \vec E^2 = \vec B^2.
\ee 
From the map to the fluid, the energy density $\rho$ is identified with $T_{00}= \frac12(E^2+B^2)$, which gives
\be
\rho = \frac{16 \left((t-z)^2+x^2+y^2+1\right)^2}{\left(t^4-2 t^2
   \left(x^2+y^2+z^2-1\right)+\left(x^2+y^2+z^2+1\right)^2\right)^3}.
\ee
One can view the dependence of the density on the space-time coordinates in different sections. For instance, in Fig.\ref{energydensity2}, 
it is drawn as a function of $t$ and $r$ at $z=0$.

Next, we identify the momentum density of the electromagnetic field with the one of the fluid, 
\be
p_i\equiv \rho v_i = T_{0i} \leftrightarrow [\vec E\times \vec B]_i\;,
\ee
which gives
\bea
p_x& =& \frac{32 \left((t-z)^2+x^2+y^2+1\right) (t x-x z+y)}{\left(t^4-2t^2
   \left(x^2+y^2+z^2-1\right)+\left(x^2+y^2+z^2+1\right)^2\right)^3},\cr
p_y&=&   -\frac{32 \left((t-z)^2+x^2+y^2+1\right) (y (z-t)+x)}{\left(t^4-2 t^2
   \left(x^2+y^2+z^2-1\right)+\left(x^2+y^2+z^2+1\right)^2\right)^3},\cr
p_z&=&   \frac{16 \left(-(t-z)^2+x^2+y^2-1\right)
   \left((t-z)^2+x^2+y^2+1\right)}{\left(t^4-2 t^2
   \left(x^2+y^2+z^2-1\right)+\left(x^2+y^2+z^2+1\right)^2\right)^3}
\eea
The velocity is found as the ratio of the momenta of the field and the energy density, giving the simple result
\bea
v_x&=&\frac{2(y+x(t-z))}{1+x^2+y^2+(t-z)^2},\cr 
v_y&=&\frac{-2(x-y(t-z))}{1+x^2+y^2+(t-z)^2},\cr
v_z^2&=&1-v_x^2-v_y^2.\label{velhopf}
\eea

\begin{figure}[t!]
\begin{tabular} {c}
\includegraphics[width=7cm]{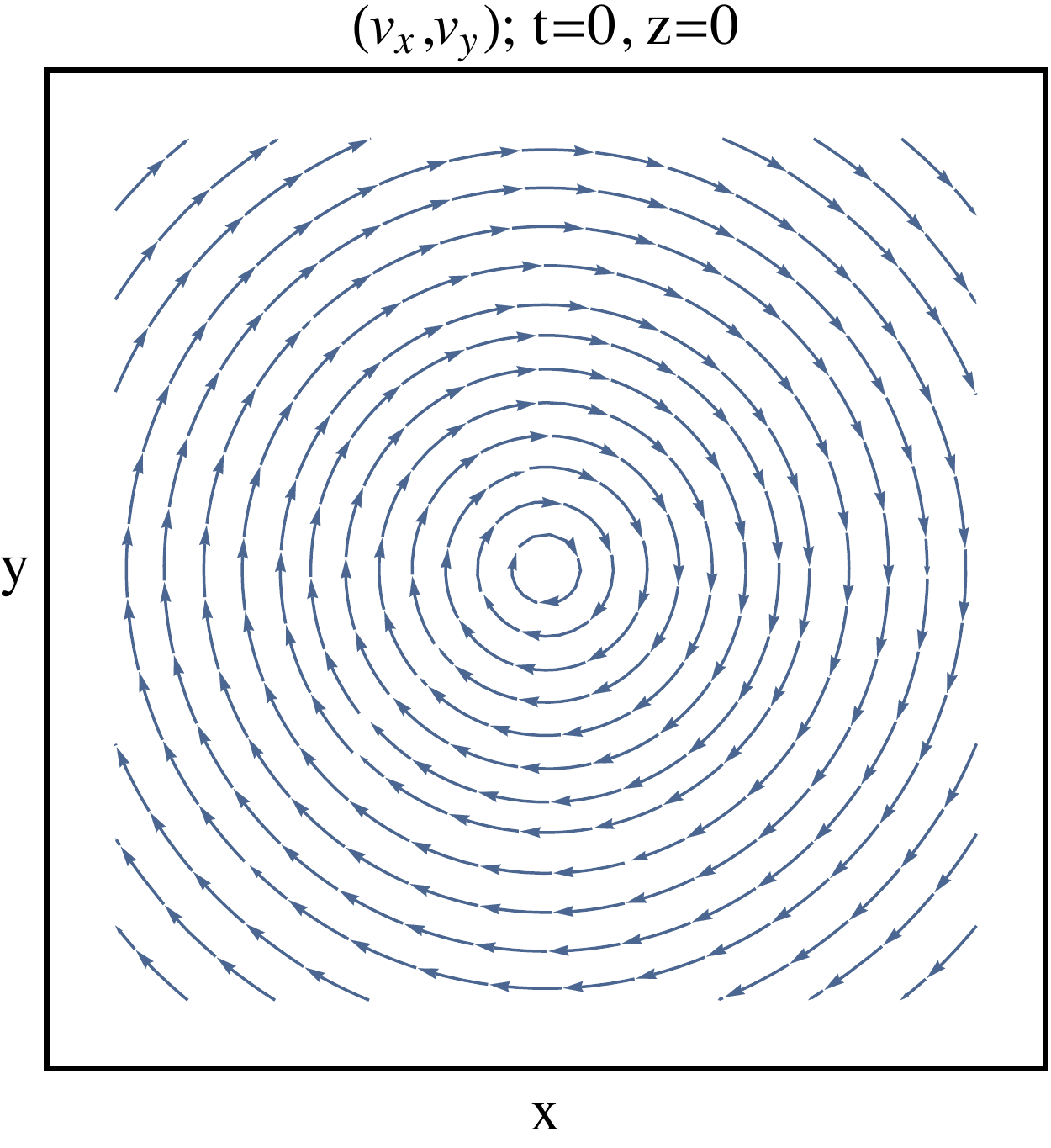}  \includegraphics[width=7cm]{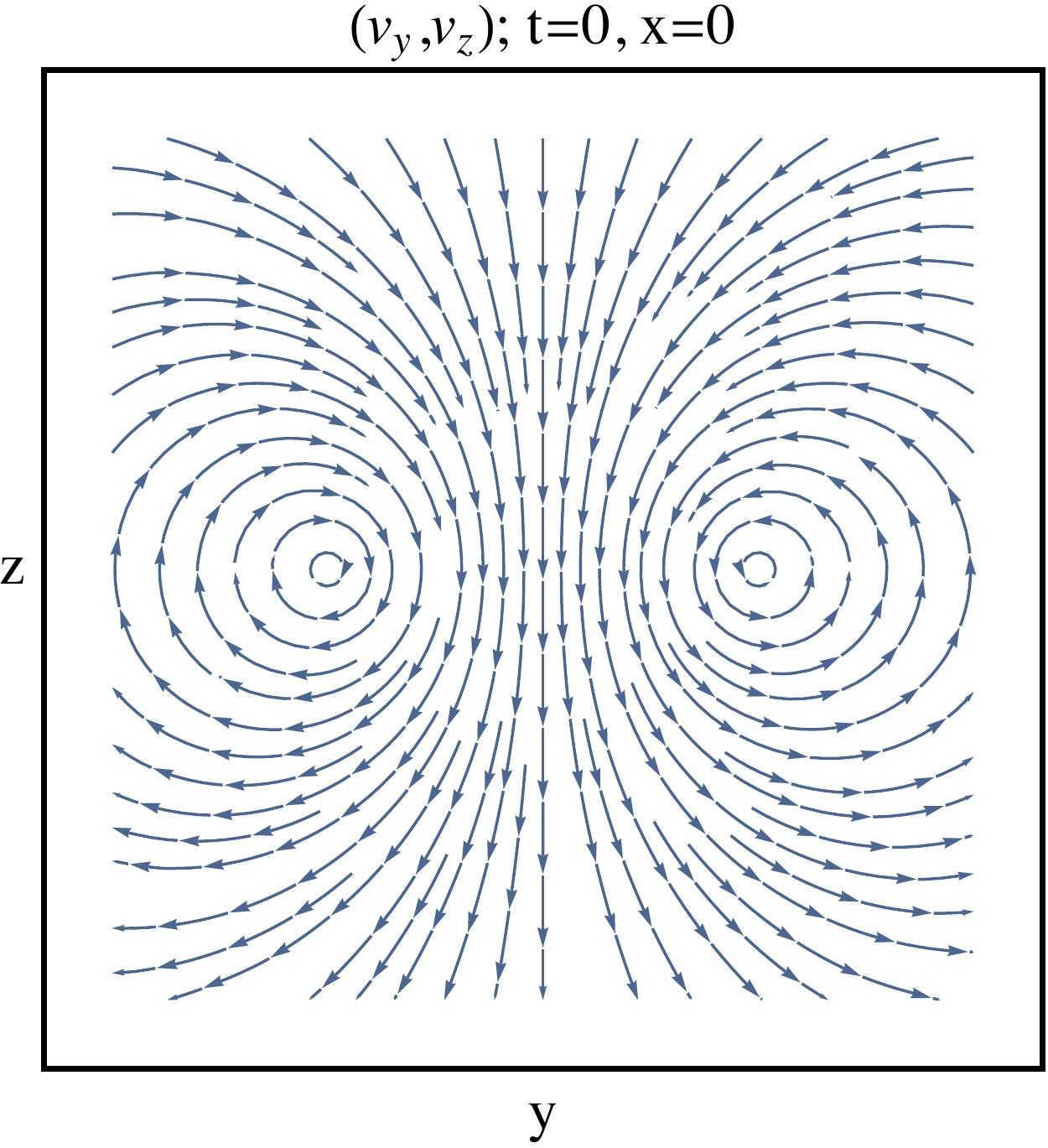}
\end{tabular}
\caption{Orthogonal sections of the velocity field for the Hopfion solution, on the $(x,y)$ plane (top) and $(y,z)$ plane (bottom).  
Using rotational symmetry in the $(x,y)$ directions the linked torus structure is apparent.}
\label{fighopf}
\end{figure}

\Comment{\bea
v_x &=& \frac{1}{\left(t^4-2 t^2
   \left(x^2+y^2+z^2-1\right)+\left(x^2+y^2+z^2+1\right)^2\right)^3}\times\cr
&&\times\frac{64 \left((t-z)^2+x^2+y^2+1\right) (t x-x z+y)
   \left(\frac{1}{4}
   \left(-t^2+x^2+y^2+z^2+1\right)^2+t^2\right)^3}{
   \left(2 t^3 (t-x y-z)-\frac{3}{2} t^2 \left((t-z)^2-x^2+y^2-1\right)
   \left(-t^2+x^2+y^2+z^2+1\right)+\frac{3}{2} t (-t+x y+z)
   \left(-t^2+x^2+y^2+z^2+1\right)^2+\frac{1}{8}
   \left((t-z)^2-x^2+y^2-1\right)
   \left(-t^2+x^2+y^2+z^2+1\right)^3\right)},\cr
v_y&=&  -\frac{1}{\left(t^4-2 t^2
   \left(x^2+y^2+z^2-1\right)+\left(x^2+y^2+z^2+1\right)^2\right)^3}\times\cr
   &&\times\frac{64
   \left((t-z)^2+x^2+y^2+1\right) (y (z-t)+x) \left(\frac{1}{4}
   \left(-t^2+x^2+y^2+z^2+1\right)^2+t^2\right)^3}{
   \left(t^3 \left((t-z)^2+x^2-y^2-1\right)+3 t^2 (t+x y-z)
   \left(-t^2+x^2+y^2+z^2+1\right)-\frac{3}{4} t
   \left((t-z)^2+x^2-y^2-1\right)
   \left(-t^2+x^2+y^2+z^2+1\right)^2-\frac{1}{4} (t+x y-z)
   \left(-t^2+x^2+y^2+z^2+1\right)^3\right)},\cr
v_z&=&\frac{1}{\left(t^4-2 t^2
   \left(x^2+y^2+z^2-1\right)+\left(x^2+y^2+z^2+1\right)^2\right)^3}\times\cr
   &&\times\frac{128
   \left(-(t-z)^2+x^2+y^2-1\right) \left((t-z)^2+x^2+y^2+1\right)
   \left(\frac{1}{4}
   \left(-t^2+x^2+y^2+z^2+1\right)^2+t^2\right)^3}{
   \left(8 t^3 (y (t-z)+x)+12 t^2 (-t x+x z+y)
   \left(-t^2+x^2+y^2+z^2+1\right)-6 t (y (t-z)+x)
   \left(-t^2+x^2+y^2+z^2+1\right)^2+(t x-x z-y)
   \left(-t^2+x^2+y^2+z^2+1\right)^3\right)}
\eea}
The last relation is due to $\vec v^2=1$. The topological structure of the solution for the velocity is described in Fig.\ref{fighopf}.

The Riemann-Silberstein vector in Bateman's construction is
\be
F^i=E^i+i B^i =\epsilon^{ijk}\partial_j\alpha\partial_k\beta\,
\ee
which means that defining $\bar{F}_i$ as the complex conjugate of $F_i$, $\vec{E}^2+\vec{B}^2=F_l\bar F_l$ and 
$2\epsilon_{ijk}E^j B^k=i\epsilon_{ijk}F^j\bar F^k$. 

Then the  velocity of the null fluid is
\begin{equation}
v_i=\frac{\epsilon_{ijk}E_jB_j}{\frac{1}{2}(\vec{E}^2+\vec{B}^2)}=\frac{\epsilon_{ijk}F^j\bar{F}^k}{F_l \bar{F}^l}.
\end{equation}

The $(p,q)$ knots and more general null solutions can be 
constructed form the Hopfion solution $\alpha_H$, $\beta_H$ changing
\begin{equation}
\alpha\to f(\alpha_H),\ \ \beta\to g(\beta_H).
\end{equation}
Then the RS vector changes by an overall factor 
\begin{equation}
F^i\to f'(\alpha_H) g'(\beta_H) F_{H}^i.
\end{equation}
This factor cancels out in the formula of the velocity. Therefore, the velocity of $(p,q)$ knots or any other 
solution constructed this way is the same as for the Hopfion, so we find no new knots. However, the energy density is different, so 
we obtain null solutions with the same velocity, but different densities.

\subsection{Some properties of the null fluid solutions}

The explicit formula for the velocity of the Hopfion is given by (\ref{velhopf}).

For $t-z=0$, the $v^a$, $a=x,y$ components of the velocity of the Hopfion are
\be\label{eq:steadyflow}
v_x=\frac{2y}{1+x^2+y^2},\ \ v_y=\frac{-2x}{1+x^2+y^2}.
\ee
Note that $\partial_a v^a=0$, since
\be
v^a=\epsilon^{ab}\partial_b\psi, \ \ \psi=\log(1+x^2+y^2).
\ee
(Note that for the usual 2+1 dimensional  Abrikosov-Nielsen-Olesen (ANO) 
vortex one has $\epsilon^{ab}\d_a \d_b \ln \phi=2\pi \delta^2(r)$, but we have $\phi\sim r$. Here 
the singularity is avoided because of the 1 inside the logarithm.)
For $t-z\neq 0$, the velocity in the $(x,y)$ plane is 
\be
v^a=\epsilon^{ab}\partial_b \psi+(t-z) \partial^a\psi,\ \ \psi=\log(1+x^2+y^2+(t-z)^2).
\ee
Define the lightcone coordinates
\be
x^\pm=t\pm z, \ \partial_\pm=\frac{1}{2}(\partial_t\pm\partial_z).
\ee
Then $x^-=x_+$ is lightcone ``time" and $x^+=x_-$ is a ``spatial" direction, and from the explicit expression above, 
we see that the velocity is constant in the $x^+$ direction, $\partial_+ v_i=0$.

 The hydrodynamic equations for the null pressureless fluid with $P=0$ and velocity $u^\mu=(1,v^i)$ are
\be
0=\partial_\mu T^{\mu\nu}=\partial_\mu(\rho u^\mu u^\nu),
\ee
and projecting onto $n_\nu=(1,0)$, we find 
\be
\partial_\mu(\rho u^\mu)=0\ \Rightarrow \ \partial_t\rho +\partial_i(\rho v^i)=0\;,
\ee
which is the continuity equation, giving the time evolution of $\rho$. 

The remaining equations for $v_i$ are the Euler equations, 
\be
u^\alpha\partial_\alpha u^\mu=0\ \Rightarrow \ \left(\partial_t+v^k\partial_k\right)v^i=0\;,
\ee
as we already explained in section 6. 

In terms of the $x^\pm$ coordinates, the Euler equations for the $v^a$, $a=x,y$ components are
\be
(1+v^z)\partial_+v^a+(1-v^z)\partial_-v^a+v^b\partial_b v^a=0\;
\ee
while the equation for $v^z$ is dependent. Since $\partial_+ v^a=0$ for the Hopfion solution, the above equations become, on the solution,
\be\label{eq:nullred}
(1-v^z)\partial_-v^a+v^b\partial_b v^a=0.
\ee

These are the Euler equations in the $t,x,y,z$ coordinates. However, we note
that they are also formally the same as the Euler equations {\em with $x^-$ as the time direction}, except for the term proportional to $v^z$. 
However, if we define $\beta^a=v^a/(1-v^z)$, the above equations become
\be
\partial_-v^a+\beta^b\partial_b v^a=0.\label{betaa}
\ee
From them, we obtain
\be
0=\epsilon^{ab}\partial_a \left(\partial_-v_b+\beta^c\partial_c v_b\right)=\partial_-\omega+\partial_a\left(\epsilon^{ab}\beta^c\partial_c v_b\right)\;,
\ee
where $\omega$ is the vorticity in the transverse directions,
\be
\omega=\epsilon^{ab}\partial_a v_b=\nabla^2 \psi,
\ee
Therefore, $\partial_-\omega$ is a total derivative and the integral will vanish if the surface term goes to zero. On the Hopfion solution,
the transverse integral of the vorticity is constant,
\be
\int d^2 x \,\omega=\lim_{\rho\to \infty} \int_0^{2\pi} d\varphi \, \rho\partial_\rho\psi=4\pi.
\ee

\subsubsection{Map to solution of the Euler equations with pressure}

For $t-z=0$, the Hopfion solution (\ref{velhopf}) can be mapped to a solution of 
the Euler equations, for a steady flow with velocity \eqref{eq:steadyflow}, constant density $\rho=1$, and pressure (note 
that we need to have $v_z=-\sqrt{1-v_x^2-v_y^2}$)
\be
\partial^a p\equiv (1-v^z)\partial_-v^a=\left(1+\sqrt{1-v_x^2-v_y^2}\right)\partial_-v^a.\label{Pmap}
\ee
Indeed, one then has 
\be
\d^a p=-v^b\partial_b v^a\;,
\ee
i.e., the Euler equation for a steady flow ($\d_t \vec{v}=0$). We consider $\rho$ constant (incompressible fluid), so the continuity equation 
{\em in 2+1 dimensions} is 
$\vec{\nabla}\cdot \vec{v}=\d_a v^a=0$.  As we saw before, for the Hopfion $\d_a v^a=0$, 
so we have a solution for an ideal fluid in 2+1 dimensions.  
One finds explicitly for the Hopfion
\be
\partial^x p=\frac{4x}{(1+x^2+y^2)^2},\ \ \partial^y p= \frac{4y}{(1+x^2+y^2)^2}.
\ee
This can be integrated to find the pressure
\be
p=p_\infty-\frac{2}{1+x^2+y^2}.
\ee
Together with \eqref{eq:steadyflow}, this describes a smooth vortex solution of the incompressible Euler equation in 2 spatial dimensions. 

The map we have found is a map from the steady state ($\d_t=0$) Euler equations with pressure (\ref{Pmap}) 
in 2+1 dimensions and the null Euler equations with $P=0$ 
in 3+1 dimensions at $t-z=0$ (also with the initial condition $\d_- v^a(t-z=0)$) and $\d_+ v^a=0$.  

We can use it in both directions: we can find solutions in 2+1 
dimensions from solutions to the null $P=0$ fluid in 3+1 dimensions, as above.

Or we can use steady flow solutions in 2+1 dimensions $(v_S^a,p_S)$ with $v_S^2<1$, as initial conditions for the null $P=0$ flows, by
\be
v^a(x^-=0)=v_S^a,\ \ \partial_-v^a(x^-=0)=\frac{\partial^a p_S}{1-v_S^2}.
\ee 
In that case,
we could construct topologically different flows by choosing the initial conditions for flows with different values of the integrated vorticity. 
It may be that some of those flows correspond to null solutions of Maxwell's equations with different topology.

\subsubsection{Reduction along $x^+$}

Another possibillity for a map to 2+1 dimensions is to define the new 2+1 dimensional velocities as
\be
\beta^a=\frac{v^a}{1-v^z}\;,
\ee
and then, since 
\be
\d_-\b^a+\b^b\d_b \b^a=\frac{\d_- v^a+\b^b \d_\b v^a}{1-v^z}+\frac{v^a}{(1-v^z)^2}\frac{v_c[\d_- v^c +\b^b\d_b v^c]}{\sqrt{1-v^dv_d}}\;,
\ee
the equations of motion of the null $P=0$ fluid  become the Euler equations  for a 2+1 dimensional fluid with constant pressure ($\vec{\nabla}P=0$) 
but in the $\b^a$ variables,
\be
\partial_-\beta^a+ \beta^b\partial_b\beta^a=0.\label{betabeta}
\ee

To this equation, we should add the continuity equation for $\rho$ in these variables,
\be
\d_- \rho+\d_a (\rho \b^a)=0.\label{contbeta}
\ee

From the Hopfion solution we get the 2+1 dimensional solution
\be
\beta^a=\epsilon^{ab}\partial_b \tilde{\psi}+(t-z)\partial^a\tilde{\psi},\ \ \tilde{\psi}=\log(x^2+y^2-1-(t-z)^2).
\ee
Note that this is a singular flow, it describes something that looks like a bouncing shock solution. 
Constant pressure implies this is a compressible flow (in fact the most compressible, since the speed of sound vanishes). 
Still the {\em initial value} for this solution satisfies the incompressibility condition
\be
\partial_a\beta^a(x^-=0)=0.
\ee
At different times $x^-$, we should use (\ref{contbeta}).

We can think of \eqref{eq:nullred} or the equivalent (\ref{betabeta}) as describing a non-relativistic fluid in $2+1$ dimensions, since the solutions are 
independent of $x^+$. In looking for new solutions we can preserve this property if we only allow the following transformations:
\begin{itemize}
\item Translations in any direction
\item Rotations in the $(x,y)$ plane
\item Boosts along the $z$ direction. For a boost
$$
t\to \cosh\tau t+\sinh \tau z, \ z\to \cosh \tau z+\sinh \tau t, 
$$
this is as time scale transformation $(t-z)\to e^{-\tau}(t-z)$.
\item Dilatations
\item Galilean boosts. These can be done as a combination of boosts and rotations. 
If we do a boost in the $x$ direction of parameter $\tau$ and then a rotation of angle $\theta$ in the $(x,z)$ plane with
$$
\sin\theta=-\tanh\tau,\ \cos\theta=\frac{1}{\cosh\tau},
$$
then
$$
t-z\to \frac{1}{\cosh\tau}(t-z),\ \ x\to x+\tanh\tau(t-z).
$$
A  further boost in the $z$ direction to remove the factor in $t-z$ will give a Galilean boost of velocity parameter $v=\sinh\tau$.
\item Special conformal transformations determined by a vector with $b_t=b_z=b$, $b_x=b_y=0$.
\end{itemize}
The last have the form of a non-relativistic special conformal transformation in the $(x,y,t-z)$ directions, if $t-z$ is identified with time 
\be
x^a \to \frac{x^a}{1-2b(t-z)}, \ t \to \frac{t+b(-t^2+r^2)}{1-2b(t-z)},\ z\to \frac{z+b(-t^2+r^2)}{1-2 b(t-z)}.
\ee
All together they form the Schr\"odinger group of non-relativistic conformal transformations.

Although they might be generalized to complex transformations, if one evaluates the velocity at the value of $t-z$ where $\partial_a v^a =0$, 
one seems to find always the same vortex solution up to real transformations (translations, rotations in $(x,y)$ or dilatations). 
This we have checked for a general special conformal transformation and for the rest of transformations with imaginary parameters.

\subsection{New solutions of Ranada type}

We can use the reverse map, from fluids to electromagnetism, to map {\em non-null } electromagnetic solutions back into fluid dynamics. 
As we saw in section 4, we can use the map between fluids and electromagnetism at $t=0$, and then use the Euler equation to evolve in time.
Also, as we saw in section 3, the procedure of giving a knotted boundary condition at $t=0$ and evolving it in time was used before for an 
incompressible fluid. 

But in fact, the condition of incompressible fluid $\vec{\nabla}\cdot \vec{v}=0$, is actually not needed. All we need is a barotropic fluid, i.e. 
with $P=P(\rho)$, in which case the helicity $H$ is still conserved. The incompressibility was used in \cite{kuznetsov1980topological} in order to 
have a nicer time evolution (to have a Hamiltonian structure with respect to the Clebsch variables $\lambda,\mu$), but that is not necessary. 

One can use the map to electromagnetism to choose a solution of Ranada type and map it to fluid at $t=0$, for instance 
the map from $\vec{v}$ 
to $\vec{A}$, and write 
\be
\vec{v}(\vec{r},t=0)=\frac{\sqrt{a}}{2\pi}\frac{4Z^2+(R^2-1)^2}{(R^2+1)^2}\times\vec{\nabla}[n\; arg(X+iY)-m\; arg(Z+i(R^2-1)/2)]
+\vec{\nabla}\chi\;,
\ee
where $\chi$ again must satisfy (\ref{chicond}). 

But to evolve to $t\neq 0$, we must use the Euler equation instead of Maxwell's equations $d*F=0$. 
Assuming as we did before that $h=$const., the Euler equation is simply
\be
\d_t\vec{v}=-(\vec{v}\cdot\vec{\nabla})\vec{v}\;,
\ee
and as we see is nonlinear, therefore hard to solve other that numerically.

In terms of the vorticity (the analog of the magnetic field), the time evolution equation is the vorticity equation (coming from the curl of the 
Euler equation)
\be
\d_t\vec{\omega}=\vec{\nabla}\times (\vec{v}\times \vec{\omega}).
\ee

Another possible knotted boundary condition at $t=0$ is a regular solution, obtained as follows. 

Take the space coordinates of the three-dimensional plane $x_i=\{x,y,z\}$, with $r^2=x^2+y^2+z^2$ and identify them with stereographic coordinates of the $S^3$
\be
X_i=\frac{2 x_i}{1+r^2}, \ \ X_4=\frac{1-r^2}{1+r^2}.
\ee
Construct the complex coordinates
\be
Z_1=X_1+i X_2, \ \ Z_2=X_3+i X_4.
\ee
The $t=0$ Ranada's solution correspond to a choice of the scalar field
\be
\phi_0=\frac{Z_1}{Z_2}.
\ee
We can generalize it to a Seifert kind of construction
\be
\phi_0 =\frac{Z_1^n}{Z_2^m}, \ n,m\geq 1 ,\ \ n,m\in \mathbb{Z}.
\ee
Take the gauge potential/velocity one-form to be defined as
\be
A=\frac{1}{4\pi i (1+|\phi_0|^2)}\left(\frac{d\phi_0}{\phi_0}-\frac{d\bar\phi_0}{\bar\phi_0} \right)+d\chi.
\ee
The magnetic field/vorticity two-form is
\be
B=\frac{1}{2\pi i}\frac{d\phi_0\wedge d\phi_0^*}{ (1+|\phi_0|^2)^2}.
\ee
The function $\chi$ has to be taken in such a way that $A$ is not singular, one can check that a minimal choice is
\be
\chi_n=-\frac{n}{2\pi} \arctan \left(\frac{y}{x}\right).
\ee
Note that, in cylindrical coordinates ($x=\rho\cos\theta, y=\rho\sin\theta$, $\hat \theta_i$ defined by $\hat \theta_idx^i/\rho=d\theta$)
\be
d\chi_n = -\frac{n}{2\pi \rho}\hat{\theta}_i dx^i= -\frac{n}{2\pi }d\theta.
\ee
We introduce stereographic coordinates for a two-sphere using the $\rho$ and $z$ coordinates
\begin{equation}
Y_1=\frac{2\rho}{1+r^2}, \ \ Y_2=\frac{2z}{1+r^2},\ \ Y_3=\frac{1-r^2}{1+r^2}.
\end{equation}
We will define the complex combination $Z_3=Y_2+i Y_3$, and the scalar
\be
\varphi =\frac{Y_1^n}{Z_3^m}, \ n,m\geq 1 ,\ \ n,m\in \mathbb{Z}.
\ee
Note that $\phi_0=e^{i n\theta}\varphi$. Then, the helicity becomes
\begin{equation}
H= \int A \wedge B= -\frac{n}{2\pi}\int d\theta\wedge \frac{d\varphi \wedge d\bar\varphi}{2\pi i(1+|\varphi|^2)^2}
=-\frac{n}{4\pi}\int  \frac{2 d\varphi \wedge d\bar\varphi}{ i(1+|\varphi|^2)^2}.
\end{equation}
The term in the integrand is the volume for for a $S^2$ parametrized by $\varphi$. As a function of $(\rho,z)$, $\varphi$ is a map 
$S^2\longrightarrow S^2$ that is winding $m$ times the circle in the $(Y_2,Y_3)$ plane, so the  target $S^2$ is wrapped $m$
 times. Therefore, the result of the integral is $4\pi m$ and one gets
\begin{equation}
H= \int A \wedge B= -nm.
\end{equation}

\section{Conclusions and discussion}

In this paper we have studied knotted solutions in electromagnetism and fluid dynamics using relations between the two theories. 
We have a map from fluid dynamics to electromagnetism, $\vec{A}=\vec{v}, \phi=h$, that works at the linearized level in the Lorenz gauge, 
or simply map initial conditions. This allowed us to generate new electromagnetic solutions from fluid solutions, or more precisely from their 
initial conditions. It also allowed us to map non-null electromagnetic solutions, of Ranada type, back into fluid dynamics, giving new types 
of barotropic $P=P(\rho)$ fluid solutions. We also found that null solutions of electromagnetism are also solutions of nonlinear electromagnetism 
theories obtained for near-constant field strenghts, like Born-Infeld, or theories obtained by integrating out quantum degrees of freedom
coupled to electromagnetism, like Euler-Heisenberg. 

We have applied the general program of writing any quantum theory at large distances as a fluid to Maxwell electromagnetism, and 
found that we can write it as a null pressureless relativistic fluid ($P=0$, $u^\mu u_\mu=0$), for which we found the Hopfion solution, which 
we described in detail. Holomorphic transformations generate new solutions, but not of $(p,q)$ type, since replacing $\a,\b$ by $\a^p,\b^q$ 
gives the same velocity $\vec{v}$, but different $\rho$. 
We were able to find, based on the Hopfion solution, a dimensional reduction in a lightcone ($x^+$) direction 
of the equations of motion for the null pressureless fluid to 
a nonrelativistic Euler equation, with nonzero pressure. That allowed us to find yet other solutions, of usual fluid dynamics, but in 2+1 dimensions. 
In particular, a {\em steady flow} (time independent) vortex corresponds to the Hopfion.

The study of knotted configuraitons in electromagnetism and hydrodynamics has obviously not exhausted itself and there are still many 
open questions to address. We list few of them:
\begin{itemize}
\item
So far, the topological non-trivial configurations analyzed are solutions of the classical equations of motion. A natural question is how to 
quantize such configurations, in particular the Hopfion and its descendants in electromagnetism. An important step in that direction is to 
identify the action for which the knotted configurations are solutions of the corresponding Euler-Lagrange equations.  This is clearly 
necessary for a path integral quantization  and is useful also for a canonical one.  The  Lagrangian density formulation inlcudes 
constraints   and hence  one needs to implement the Dirac's bracket procedure for a canonical quantization. 
\item
The generalization to the DBI electromagnetic action presented in section 2.5 is based on quantities that are analogous to 
electromagnetism in non-trivial medium. 
So far, most of the discussion of the knotted solutions has been in an empty medium. It  is of great interest, in particular for the 
experimental realization of such configurations, to examine their dependence on the dielectric and magnetic permeability 
of the medium, both when the latter are constants and when they are space-time dependent. 
\item
The latter topic involves an interaction between the EM fields  and charges and currents of the medium. Just as interesting 
is the question of what will be the effect of the knotted solutions on  the motion of external charged particles and currents. This 
again is probably needed for the experimental detection of knotted EM configurations.  From the other direction,  the question 
is if there  are such  topological solutions  also to   the non-free Maxwell's equations  that include currents and charge density. 
Similarly, it should be interesting to explore hydrodynamics systems with external sources. 
\item
In  section  6  we derived a map between the knotted EM solutions and solutions of pressureless hydrodynamics. An immediate question 
is whether one can generalize this map, and in fact the null property, to also find topologically non-trivial solutions of hydrodynamics 
that include non-vanishing  pressure.
\item
In  light of the fluid/gravity  correspondence that has been suggested recently, a relevant question is if one can translate the
knotted solutions of hydrodynamics to topological non-trivial solutions of gravity. One can also approach this using various maps that have been suggested 
between gauge dynamics and gravity. It will be extremely interesting to find the gravitational analogs of Hopfion and its relatives. 
\end{itemize}
\section*{Acknowledgements}
We would like to thank Manuel Array\'as, Ori Ganor and Nilanjan Sircar for useful comments and discussions.
HN is grateful for the hospitality of the Department of Physics at Tel Aviv University, during the final stages of this work. 
This work was supported in part by a center of excellence supported by the Israel Science Foundation (grant number 1989/14), and by 
the US-Israel bi-national fund (BSF) grant number 2012383 and the Germany Israel bi-national fund GIF grant number I-244-303.7-2013. 
The work of HN is supported in part by CNPq grant 304006/2016-5 and FAPESP grant 2014/18634-9. HN would also 
like to thank the ICTP-SAIFR for their support through FAPESP grant 2016/01343-7. C.H. is supported by the Ramon y
Cajal fellowship RYC-2012-10370, the Asturian grant FC-15-GRUPIN14-108 and the Spanish national
grant MINECO-16-FPA2015-63667-P. D.F.W.A. is supported by CNPq grant 146086/2015-5.

\bibliography{Hopfionpaper}
\bibliographystyle{utphys}

\end{document}